\documentclass{article}

\usepackage{PRIMEarxiv}

\usepackage[utf8]{inputenc} 
\usepackage[T1]{fontenc}    
\usepackage{hyperref}       
\usepackage{url}            
\usepackage{booktabs}       
\usepackage{amsfonts}       
\usepackage{nicefrac}       
\usepackage{microtype}      
\usepackage{lipsum}
\usepackage{fancyhdr}       
\usepackage{graphicx}       
\graphicspath{{media/}}     

\usepackage{booktabs}
\usepackage{multirow}
\usepackage{amsfonts}
\usepackage{makecell}
\usepackage{threeparttable}
\usepackage{xr}
\usepackage{amsmath}
\usepackage{siunitx}

\title{Learning light scattering from operator parameter spaces to Galerkin-consistent solution spaces}

\author{
  Lida Liu \\
  School of Optical and Electronic Information\\
  Huazhong University of Science and Technology \\
  Wuhan\\
   \And
  Jingwei Wang \\
  School of Electronic Information and Communications\\
  Huazhong University of Science and Technology\\
  Wuhan
  \And
  Wei Cao, Yang Zhang \\
  School of Optical and Electronic Information\\
  Huazhong University of Science and Technology\\
  Wuhan
  \And
  Yuntian Chen \\
  School of Optical and Electronic Information\\
  Wuhan National Laboratory of Optoelectronics\\
  Huazhong University of Science and Technology\\
  Optics Valley Laboratory\\
  Wuhan \\
  \texttt{yuntian@hust.edu.cn}
}

\begin{document}
\maketitle

\begin{abstract}
Efficient and generalizable full-wave simulation is essential for nanophotonic analysis and inverse design, yet existing methods face a tradeoff between the high computational cost of numerical solvers and the limited generalizability of neural operator models for complex optical scattering. Here, we introduce FEMONet, a finite-element-constrained operator-learning framework that learns light scattering from an operator parameter space to a Galerkin-consistent solution space. The operator parameter space encodes the physical entities defining a wave-equation problem, while the variational weak form links this space to the coordinate and physical solution spaces. Integrated with operator-learning networks, FEMONet extends classical solvers from isolated problem instances to parameterized scattering operators. To our knowledge, FEMONet represents the first Galerkin-consistent operator-learning framework for complex-valued optical scattering, grounded in the variational weak form of the governing vector wave equations. Finite-element discretization absorbs spatial derivatives into assembled stiffness matrices and load vectors, removing coordinate-based derivatives of the neural-network output from the physics loss and improving training efficiency. By predicting finite-element expansion coefficients rather than unconstrained field values, the Galerkin-consistent formulation preserves compatible trial and test spaces, achieving high accuracy, stable training, and generalization across dielectric, metallic, arrayed, plasmonic, and three-dimensional nanophotonic structures.
\end{abstract}

\keywords{operator parameter space \and operator learning \and finite elements method \and variational principle}

\section{Introduction}
In nanophotonics, numerical solvers are essential for modeling wave-scale light-matter interactions and engineering of artificial structures\cite{jin2015finite,teixeira2023finite,pelosi2009quick,harrington1993field}. These phenomena—including plasmonic field enhancement\cite{maier2007plasmonics}, QNM-based resonant scattering\cite{kristensen2014modes,lalanne2013theory}, and metasurface wavefront control\cite{yu2011light}—rely on these solvers to provide full-wave solutions to Maxwell's equations. Although traditional methods such as the Finite-Difference Time-Domain (FDTD)\cite{teixeira2023finite}, Finite Element Method (FEM)\cite{jin2015finite,pelosi2009quick}, and Method of Moments (MoM)\cite{harrington1993field} are established tools for solving Maxwell’s equations, they face severe bottlenecks in modern optimization and inverse design tasks\cite{molesky2018inverse,jiang2021deep}. Operator solvers in these multi-parameter spaces scenarios demand the repeated assembly and solving of large matrices, leading to prohibitive computational costs. Such overhead significantly hinders design efficiency and rapid prototyping, creating an urgent need for more efficient simulation paradigms to navigate increasingly complex nanophotonic devices.

To circumvent this heavy computational burden, Physics-Informed Neural Networks (PINNs) have emerged as a promising paradigm for solving partial differential equations (PDEs)\cite{raissi2019physics}. By encoding physical laws into the loss function, PINNs can approximate solutions to Maxwell's equations without requiring labeled data\cite{11chen2020physics,12lim2022maxwellnet}. Meanwhile, driven by the pursuit of better numerical stability and broader generalization, the landscape of physics-informed learning has expanded rapidly. On the one hand, methods such as Variational PINNs (VPINNs) and hp-VPINNs bridge PINNs with classical variational principles, establishing the loss function by integrating the PDEs residual multiplied by a set of basis functions\cite{VPINN,hpVPINN}. This reduces the order of the differential operator, which can effectively lower the required regularity in the output of the neural network (NN). On the other hand, a more profound shift has occurred from point-to-value learning to infinite-dimensional function-to-function mapping, known as operator learning. Notable frameworks include DeepONet\cite{15lu2021learning}, Physics-Informed DeepONet\cite{16wang2021learning} and Physics-Informed Fourier Neural Operator\cite{17li2020fourier}.

Despite these advances, three fundamental barriers still prevent physics-informed and operator-learning methods from becoming reliable simulation tools for practical nanophotonics. First, many PINN-based methods remain demonstrated mainly on canonical benchmarks or problem-specific simplified structures \cite{raissi2019physics,11chen2020physics,12lim2022maxwellnet}. The optimization of strong-form PDE residuals is sensitive to loss balancing, hyperparameters, material discontinuities, and high-index or metallic losses, which makes their transfer to realistic plasmonic, metasurface, and multi-scatterer systems challenging \cite{18WangUndersan,19wang2022and,jiang2021deep,20ma2021deep,21wiecha2021deep}. Second, although neural operators provide a powerful framework for learning mappings between function spaces and solution operators of parametric PDEs \cite{15lu2021learning,16wang2021learning,17li2020fourier,Kovachki2023NeuralOperator}, their generalization in optical scattering is often difficult to interpret physically. A scattering solution operator is not determined by geometry alone, but by the physical entities that define the wave-equation problem, including wavelength, material distribution, excitation, and boundary conditions. Previous work on photonic crystals has shown that encoding Maxwell-operator structure can enable machine prediction of topological transitions beyond simple parameter interpolation \cite{Wu2020MachinePrediction}, suggesting the importance of physically structured operator representations. However, a general formulation that defines the physically sufficient domain and boundary of neural-operator generalization for full-wave optical scattering remains lacking. Third, from a numerical-analysis perspective, weak-form neural solvers such as VPINNs and hp-VPINNs are typically formulated in a Petrov-Galerkin setting, where the neural-network trial space and the prescribed test space are constructed separately \cite{VPINN,hpVPINN}. Although this formulation reduces the differentiation order of the residual, it does not by itself enforce the trial-test compatibility that underlies Galerkin finite-element discretizations or related stability conditions such as the inf-sup condition \cite{ErnGuermond2004,BoffiBrezziFortin2013,Bathe2001,Rojas2024RVPINN}. These limitations call for a neural-operator framework that is simultaneously applicable to realistic optical structures, physically interpretable in its generalization domain, and consistent with the discrete variational structure of classical full-wave solvers.

To address these challenges, we propose FEMONet, an operator-parameter-space-augmented FEM-constrained operator-learning framework for complex-valued full-wave optical scattering. The central idea is to reformulate optical simulation from solving isolated boundary-value problems into learning a family of wave-equation solution operators defined over an operator parameter space. This space is not merely a larger collection of input variables; it provides a physically sufficient domain on which neural-operator generalization is meaningful, interpretable, and bounded. Each operator-parameter instance is defined by the physical entities that determine the governing differential operator, source term, and boundary condition. Conventional geometry or design parameters can be regarded as local coordinates only when they uniquely determine this operator-defining physical representation. In contrast, representing scatterers through material distributions, together with wavelength, excitation, and invariant physical settings, places diverse geometries and optical conditions within a shared scattering-operator family. FEMONet then connects this operator parameter space to the physical solution space through the variational weak form and FEM discretization. Rather than predicting unconstrained field values, the network predicts finite-element expansion coefficients and is trained by a preconditioned FEM residual assembled from the corresponding stiffness matrix and load vector. This Galerkin-consistent construction preserves the compatible trial-space structure of the underlying FEM discretization, avoids coordinate-based automatic differentiation of the neural-network output, and transfers the weak-form structure of FEM from single-instance solvers to parameterized neural operators. We validate this framework on practical optical scattering problems, including dielectric and metallic scatterers, coupled multi-scatterer configurations, plasmonic nanostructures, and three-dimensional metasurfaces, demonstrating accurate, efficient, and physically bounded generalization across prescribed operator-parameter spaces.

\section{Results}

\subsection{FEMONet learns optical scattering over an operator parameter space}

Taking the two-dimensional optical scattering problem as an example, under transverse magnetic (TM) polarization, the scattered electric-field component $E_z^s$ satisfies a wave equation that can be written in differential-operator form as \cite{jin2015finite}
\begin{equation}
\label{eqn3}
\mathcal{L}\phi=f
\quad
\text{in } \Omega, 
\end{equation}
where 
$\mathcal{L}=\nabla\times\left(\overline{\mu}_r^{-1}\nabla\times\right)-k_0^2\overline{\varepsilon}_r$, $\phi=E_z^s(\boldsymbol{r})$, and $f=-\nabla\times\overline{\mu}_r^{-1}\nabla\times \hat{\boldsymbol{z}}E_z^i(\boldsymbol{r})+k_0^2\overline{\varepsilon}_r\hat{\boldsymbol{z}}E_z^i(\boldsymbol{r}).$
In this work, two related but distinct operators are involved: the wave-equation differential operator $\mathcal{L}$, which governs each boundary-value problem, and the scattering solution operator $\mathcal{G}_s$, which maps the physical entities defining the problem to the scattered field evaluated at a spatial coordinate:
\begin{equation}
\label{so}
E_z^s(\boldsymbol{r})=\mathcal{G}_s(\lambda,\varepsilon_r,\mu_r,E^i,\mathcal{B})(\boldsymbol{r}). 
\end{equation}

We define the optical scattering problem in an operator parameter space $\mathcal{P}$, where each point corresponds to a specific wave-equation problem instance,
\begin{equation}
p=(\lambda, \varepsilon_r, \mu_r, E^i, \mathcal{B})\in\mathcal{P}. 
\end{equation}
Here, $\lambda$, $\varepsilon_r$, $\mu_r$, $E^i$, and $\mathcal{B}$ denote the wavelength, relative permittivity, relative permeability, incident field, and boundary conditions, respectively. These physical entities jointly determine the parameterized differential operator $\mathcal{L}_p$, the source term $f_p$, and the boundary conditions. Therefore, optical scattering is not treated as a set of isolated forward simulations, but as a family of parameterized boundary-value problems,
\begin{equation}
\label{boundary-value_porblems}
\mathcal{L}_p \phi_p = f_p \quad \text{in } \Omega, \quad
\mathcal{B}_p\phi_p = g_p \quad \text{on } \partial\Omega, 
\quad p \in \mathcal{P}, 
\end{equation}
where $\phi_p$ denotes the scattered-field solution associated with the problem instance $p$. The objective of FEMONet is to learn the parameterized scattering solution operator over $\mathcal{P}$, rather than repeatedly solving each individual operator instance.

The variational weak form and FEM discretization provide the bridge between the operator parameter space, the coordinate space, and the physical solution space. For each $p\in\mathcal{P}$, the physical parameters are encoded into a stiffness matrix $A_p$ and a load vector $b_p$, while the spatial dependence of the field is represented by FEM basis functions, $E_z^s(\boldsymbol{r})\approx\sum_{i=1}^{N}u_iN_i(\boldsymbol{r}).$ The corresponding discrete system is
\begin{equation}
\label{Ab}
A_pu_p=b_p. 
\end{equation}
The elements in $A_p$ given by
\begin{equation*}
      A_{ij}
      =\int_\Omega[\nabla\times\hat{z}N_i \cdot \overline{\mu_r}^{-1}\nabla\times \hat{z}N_j-k_0^2\overline{\varepsilon}_r\hat{z}N_i\cdot\hat{z}N_j]dS 
      +\int_{\delta\Omega}[\hat{z}N_i\cdot jk_0\boldsymbol{n}\times\hat{z}N_j\times\boldsymbol{n} ]dl,
\end{equation*}
and the elements in $b$ given by
\begin{equation*}
        b_i
        =\int_{\delta\Omega}[\hat{z}N_i\cdot\boldsymbol{n}\times\nabla\times\hat{z}E_z^i]dl-\int_\Omega[\nabla\times\hat{z}N_i\cdot\overline{\mu_r}^{-1}\nabla\times\hat{z}E_z^i-k_0^2\overline{\varepsilon}_r\hat{z}N_i\cdot\hat{z}E_z^i]dS.
\end{equation*}
In classical FEM, this linear system is assembled and solved independently for each $p$. We have included the complete variational derivation process in the SI Appendix.

In FEMONet, the neural network instead predicts the FEM expansion coefficients, $E_z^s(\boldsymbol{r})
\approx\sum_{i=1}^{N}u_{p,\theta,i}N_i(\boldsymbol{r}),$ and is trained by minimizing the FEM residual over sampled operator instances,
\begin{equation}
\label{loss}
\min_{\theta} L(\theta)=\frac{1}{N_s}\sum_{s=1}^{N_s}\frac{1}{N_{p_s}}||M_{p_s}^{-1}\left(A_{p_s}u_{p_s,\theta}-b_{p_s}\right)||^2_2, 
\end{equation}
where $\theta$ represents the network hyperparameters and $M_{p_s}^{-1}$ denotes the preconditioner used to improve the conditioning of the residual minimization, $||\cdot||_2$ represents the $L_2$ norm. This formulation combines operator learning with Galerkin-consistent physical constraints and extends classical operator solving from individual problem instances to the operator parameter space.

\begin{figure}
\centering
\includegraphics[width=17cm]{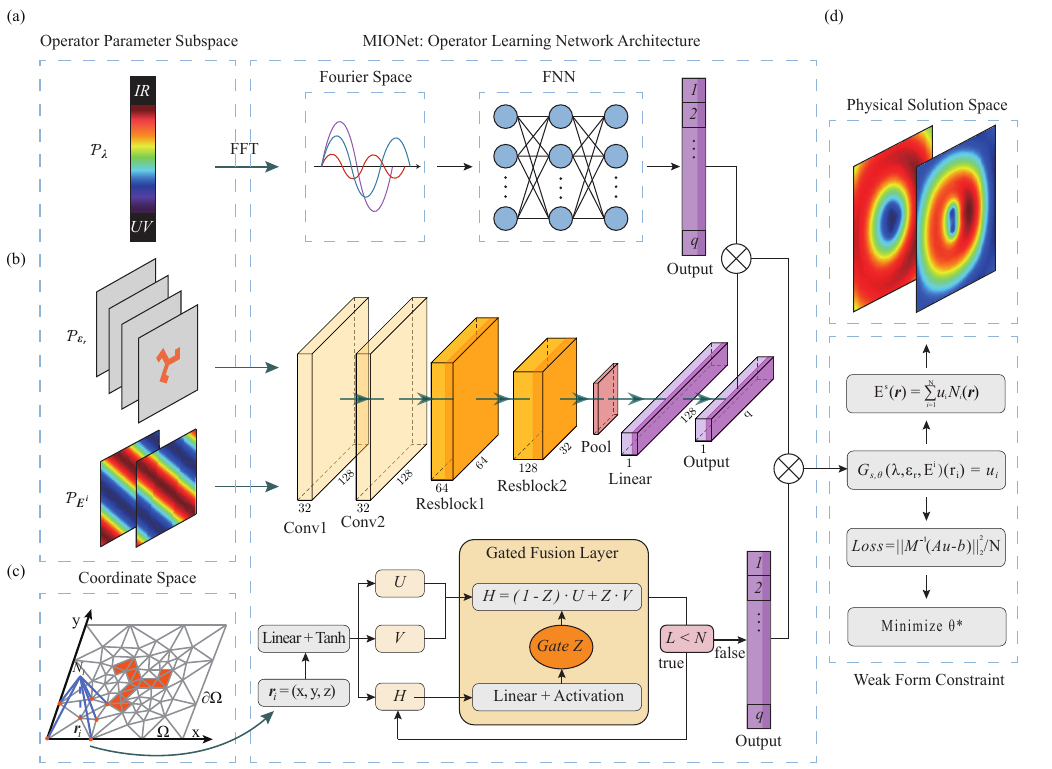}
\caption{
Operator-parameter-space-augmented MIONet architecture for optical scattering operator learning.
(a) The wavelength branch encodes the spectral subspace $\mathcal{P}_{\lambda}$ using Fourier features and a feed-forward neural network.
(b) The material and excitation branches encode the relative permittivity images $\varepsilon_r$ and incident field images $E^i$, corresponding to $\mathcal{P}_{\varepsilon_r}$ and $\mathcal{P}_{E^i}$, respectively.
(c) The trunk network encodes each spatial query point $\boldsymbol{r_i}\in\Omega$ corresponding to the basic function $N_i$ and therefore represents the coordinate space.
(d) The branch and trunk features are fused to predict the FEM expansion coefficient
$u_{p,\theta,i}=\mathcal{G}_{s,\theta}(\lambda,\varepsilon_r,E^i)(\boldsymbol{r}_i)$.
The weak-form constraint provides a physics-consistent optimization objective through the preconditioned FEM residual, and the scattered field is reconstructed in the physical solution space as
$E_z^s(\boldsymbol{r})=\sum_{i=1}^{N}u_iN_i(\boldsymbol{r})$.
}
\label{fig2}
\end{figure}

Considering the scattering solution operator defined in Eq.~(\ref{so}), we design an operator-parameter-space-augmented MIONet framework, as illustrated in Fig.~\ref{fig2}. MIONet is selected as the operator-learning architecture because its multi-input formulation is naturally compatible with the structure of the operator parameter space. Instead of representing the scattering problem by a single input function, MIONet learns a solution operator defined on the multiple input-function spaces. This formulation matches the fact that an optical scattering operator is jointly determined by several physical entities, including the wavelength, material distribution, excitation field, and invariant physical settings.

For the scattering problems considered in this work, the operator parameter space is decomposed as
\begin{equation}
\mathcal{P}
=
\mathcal{P}_{\lambda}
\times
\mathcal{P}_{\varepsilon_r}
\times
\mathcal{P}_{E^i}
\times
\mathcal{P}_{I}. 
\end{equation}
Here, $\mathcal{P}_{\lambda}$, $\mathcal{P}_{\varepsilon_r}$, and $\mathcal{P}_{E^i}$ denote the spectral, material, and excitation subspaces, respectively. The invariant subspace $\mathcal{P}_{I}$ contains fixed physical settings, such as the permeability $\mu_r$ and boundary condition $\mathcal{B}$. Since these quantities remain unchanged within the operator family considered in this work, they are not assigned to an additional branch network, but are incorporated as fixed conditions through the overall network setting and the FEM-constrained loss.

Under this decomposition, each branch network can be interpreted as an encoder for a projected subspace of $\mathcal{P}$. Specifically, the input-function spaces of the branch networks, denoted by $\mathcal{V}_{\lambda}$, $\mathcal{V}_{\varepsilon_r}$, and $\mathcal{V}_{E^i}$, provide functional representations of the projected operator-parameter subspaces $\mathcal{P}_{\lambda}$, $\mathcal{P}_{\varepsilon_r}$, and $\mathcal{P}_{E^i}$, respectively. In contrast, the trunk network takes the spatial query point $\boldsymbol{r}\in\Omega$ as input and therefore represents the coordinate space. The branch and trunk features are then fused to predict the FEM expansion coefficient, $u_{p,\theta,i}=\mathcal{G}_{s,\theta}(\lambda,\varepsilon_r,E^i)(\boldsymbol{r}_i),$
which reconstructs the scattered field in the physical solution space through the FEM basis expansion. In this way, the MIONet architecture provides a direct network realization of the proposed theoretical framework: branch networks encode projections of the operator parameter space, the trunk network encodes the coordinate space, the weak-form constraint provides a physics-consistent optimization objective through the preconditioned FEM residual and the scattered field is reconstructed in the physical solution space.

\begin{table}[t!]
\centering
\begin{threeparttable}
\caption{Operator-parameter-space map}
\label{tab:result_summary}
\begin{tabular}{c c c c}
\toprule
\makecell[c]{Case} & \makecell[c]{$\mathcal{P}_{sub}$}
& \makecell[c]{Network \\ Architecture}  & \makecell[c]{Validation \\ Dimensions}
\\
\midrule
\makecell[c]{1.Basic \\ scatterers} & $\mathcal{P}_{\varepsilon_r}$ & \makecell[c]{Branch net 2 \\ \& Trunk net} &\makecell[c]{\makecell[c]{Performance \\ comparison,}  \\ \makecell[c]{Extrapolation \\ capability}} \\
\midrule
\makecell[c]{2.Metal \\ scatterers} & $\mathcal{P}_{\lambda},\mathcal{P}_{\varepsilon_r},\mathcal{P}_{E^i}$ & \makecell[c]{Branch net 1,2,3 \\ \& Trunk net} &\makecell[c]{Architectural \\ efficiency} \\
\midrule
\makecell[c]{3.Multiple \\ metal scatterers} & $\mathcal{P}_{\varepsilon_r},\mathcal{P}_{E^i}$ & \makecell[c]{Branch net 2,3 \\ \& Trunk net} & \makecell[c]{Hyperparameter \\ sensitivity} \\
\midrule
\makecell[c]{4.Plasmonic \\ nanostructure} & $\mathcal{P}_{\lambda},\mathcal{P}_{\varepsilon_r}$ & \makecell[c]{Branch net 1,2 \\ \& Trunk net} & \makecell[c]{Sparse-sample \\ learning} \\
\midrule
\makecell[c]{5.3D metasurface} & $\mathcal{P}_{\lambda},\mathcal{P}_{\varepsilon_r}$ & \makecell[c]{Branch net 1,2 \\ \& Trunk net} & \makecell[c]{Dimensional \\ expansion} \\
\bottomrule
\end{tabular}
\begin{tablenotes}
\footnotesize
\item The listed $\mathcal{P}_{sub}$ are the active operator projections subspaces for each problem family. Omitted subspaces are invariant and are incorporated through fixed physical settings and the FEM residual, rather than being ignored.
\end{tablenotes}
\end{threeparttable}
\end{table}

We next validate this operator-parameter-space framework across five representative optical scattering problems, as summarized in Table~\ref{tab:result_summary}. These cases are organized according to the active projections of the operator parameter space, including the spectral subspace $\mathcal{P}_{\lambda}$, material subspace $\mathcal{P}_{\varepsilon_r}$, and excitation subspace $\mathcal{P}_{E^i}$. Each case activates a different combination of branch-network inputs and is designed to test a distinct aspect of FEMONet, including accuracy, efficiency, extrapolation, architectural efficiency, hyperparameter robustness, sparse-sample learning, and extension from two-dimensional scatterers to three-dimensional metasurface structures.

\subsection{Galerkin-consistent learning and bounded generalization}
\begin{figure}
    \centering
    \includegraphics[width=14cm]{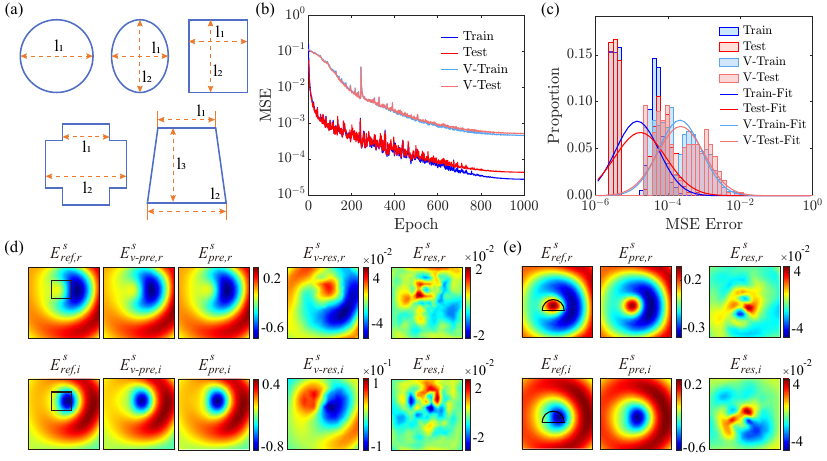}
    \caption{Comparative ablation and generalization study on basic lossless scatterers. (a) Geometric structures of the scatterers. (b) MSE convergence curves. (c) MSE histograms with corresponding Gaussian fits. (d) Scattering field distributions for representative lossless scatterer. Columns (left to right): FEM reference $E_{ref}^s$, VPINN prediction $E_{v-pre}^s$, FEMONet prediction $E_{pre}^s$, and their respective errors $E_{v-res}^s$ and $E_{res}^s$. (e) Bounded generalization within a shared operator parameter space of FEMONet evaluated on unseen scatterer geometry (left to right: FEM reference $E_{ref}^s$, FEMONet prediction $E_{pre}^s$, and errors $E_{res}^s$). Rows indicate the real and imaginary parts.}
    \label{fig3}
\end{figure}

We first evaluate FEMONet on basic lossless scatterers, for which the active projection of the operator parameter space is the material subspace $\mathcal{P}_{\varepsilon_r}$. This setting provides a controlled benchmark to isolate the effect of the FEM-constrained residual, because the wavelength, excitation field, permeability, and boundary conditions are fixed while the material geometry varies. Five representative dielectric scatterers with a refractive index of $1.45$ are placed in a $2~\mu\mathrm{m}\times2~\mu\mathrm{m}$ computational domain and excited by a TM-polarized plane wave at $\lambda=1.55~\mu\mathrm{m}$. The sample set contains $1,558$ unlabeled operator instances, divided into $1,246$ training and $312$ testing samples. 

To ensure a stringent and fair comparison, we construct an enhanced VPINN baseline, in which the scattering boundary condition is incorporated directly into the weak-form residual instead of being imposed as a separate boundary penalty. FEMONet is compared with enhanced VPINN baseline using the identical DeepONet configuration. Thus, the performance difference mainly reflects the role of the physical constraint rather than the network architecture. The efficiency gain can be understood from the differentiation order imposed on the neural network. For wave equations, PINNs minimize the strong-form residual and require second-order coordinate derivatives of the network output. VPINNs use integration by parts in the weak form to reduce this requirement to first-order derivatives\cite{raissi2019physics,VPINN,hpVPINN}. In contrast, FEMONet applies the differentiation to the basis cunctions by FEM discretization and matrix assembly which further reduce this requirement to zero-order, while the predicted coefficients remain constrained by the Galerkin-consistent FEM system. As shown in Fig.~\ref{fig3}(b,c), FEMONet converges faster and achieves an order-of-magnitude lower MSE than the enhanced VPINN. The gaussian fitting parameters for all cases in SI Appendix. On a single NVIDIA RTX 4090 GPU, FEMONet completes $1,000$ training epochs in $2~\mathrm{h}~11~\mathrm{min}~31~\mathrm{s}$, while enhanced VPINN requires $17~\mathrm{h}~27~\mathrm{min}~41~\mathrm{s}$. The field maps in Fig.~\ref{fig3}(d) further show that FEMONet reproduces both the real and imaginary parts of the scattered field with substantially smaller residuals. Furthermore, the maximum errors of FEMONet are localized within specific regions, in contrast to the globalized error distribution exhibited by VPINN. A detailed description of the enhanced VPINN baseline settings, together with an analysis of the sources of the FEMONet efficiency advantage over VPINN in SI Appendix.

We further examine whether the learned scattering operator can transfer across geometry families within the prescribed operator parameter space. From the viewpoint of conventional design parameter space, this task corresponds to generalization across unseen topological or shape classes, because different scatterer families are usually described by different local design parameters, such as $g=(r,l,w,\ldots)$. Such parameters can be effective within a fixed geometry parameter space, but they do not provide a common parameterization across different shapes and topologies. In the operator parameter spaces view, these local design parameters are lifted to an operator-defining physical entities,
\begin{equation*}
    g=(r,l,w,...)\xrightarrow{\Psi} p=(\lambda,\varepsilon_r,\mu_r,E^i,B) \xrightarrow{\mathrm{FEM}} (A_p,b_p),
\end{equation*}
where $\Psi$ represents sufficient and single-valued conditions. In this situation, the material distribution $\varepsilon_r(\mathbf r)$ uniformly represents different geometries within a operator parameter subspace $P_{\varepsilon_r}$. As shown in Fig.~\ref{fig3}(e), FEMONet accurately predicts the scattered fields for scatterer shapes that are not included in the training set. Across 320 unseen parameter instances from five additional geometries, the model achieves a test MSE of $1.4\times10^{-3}$. This result should not be interpreted as unrestricted extrapolation to arbitrary optical structures. Rather, it shows that what appears as cross-topology generalization in conventional design-parameter space can be understood, in the operator-parameter-space formulation, as transfer within a shared operator parameter space. Thus, conventional design parameters act as local coordinates when they sufficiently determine the scattering operator, whereas the operator parameter space provides the common physical space in which different coordinates, geometries, and topologies can be represented uniformly. More details of the performance on unseen samples are provided in the SI Appendix.

\subsection{Coupled parameter-subspaces learning in metallic scatterers}
\begin{figure}
    \centering
    \includegraphics[width=15cm]{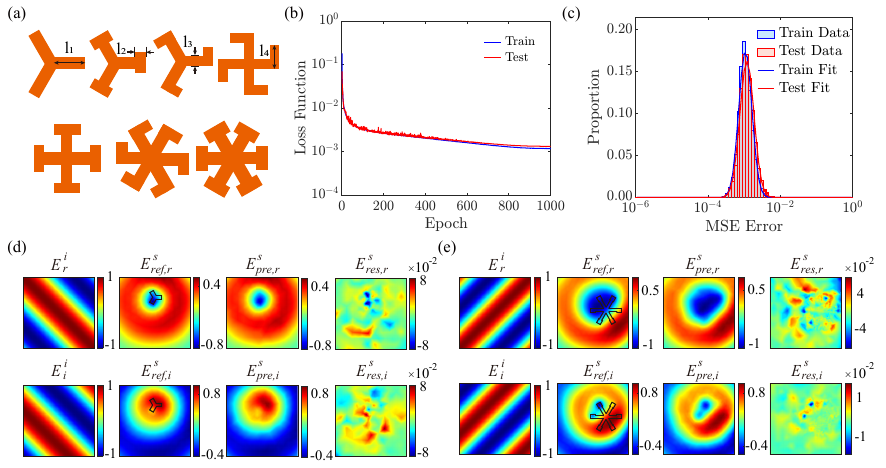}
    \caption{(a) Geometric structures of the scatterers. (b) Loss convergence curves. (c) MSE histograms with corresponding Gaussian fits. (d, e) Scattering fields for two metallic scatterers. Columns (left to right): Incident field $E^i$, FEM reference $E_{ref}^s$, FEMONet prediction $E_{pre}^s$ and error $E_{res}^s$. Rows indicate the real and imaginary parts.}
    \label{fig4}
\end{figure}
We next evaluate FEMONet on single metallic scatterers, where the scattering operator depends jointly on the spectral, material, and excitation subspaces, $\mathcal{P}_{\lambda}\times\mathcal{P}_{\varepsilon_r}\times\mathcal{P}_{E^i}$. This case tests whether the MIONet architecture can efficiently represent a higher-dimensional operator parameter space in which wavelength-dependent material dispersion, complex-valued permittivity, and incident direction are simultaneously varied.

Seven gold scatterers with characteristic lengths ranging from $320$ to $620~\mathrm{nm}$ are considered, as shown in Fig.~\ref{fig4}(a). The wavelength spans $1.5\sim2~\mu\mathrm{m}$, and the incident TM-polarized plane wave is sampled over angles from $0^\circ$ to $315^\circ$ with a $45^\circ$ step. Gold dispersion is described by the Drude model. The resulting operator-parameter sample set contains $50,688$ unlabeled instances, with $40,550$ used for training and $10,138$ used for testing.

The loss curves in Fig.~\ref{fig4}(b) show that the training and testing errors converge consistently and stabilize around $1.5\times10^{-3}$. The MSE distributions in Fig.~\ref{fig4}(c) remain closely aligned between training and testing sets, indicating that FEMONet does not overfit despite the enlarged parameter space. Representative field predictions in Fig.~\ref{fig4}(d,e) further confirm that the model accurately captures metallic scattering responses under different incident conditions.

This case also demonstrates architectural efficiency. Moving from the dielectric benchmark to the metallic scattering task increases the number of trainable parameters from $0.43$M to $2.57$M, a $5.98\times$ increase. In contrast, the number of operator-parameter configurations increases from $1,558$ to $50,688$, approximately $32.5\times$. Moreover, the input representation expands from a single real-valued material map to wavelength, complex-valued permittivity, and complex-valued incident fields. Therefore, the increase in model size is moderate relative to the expansion of the operator parameter space, supporting the scalability of the multi-branch MIONet architecture.

\subsection{Robustness under hyperparameter variation}
\begin{figure}
    \centering
    \includegraphics[width=14cm]{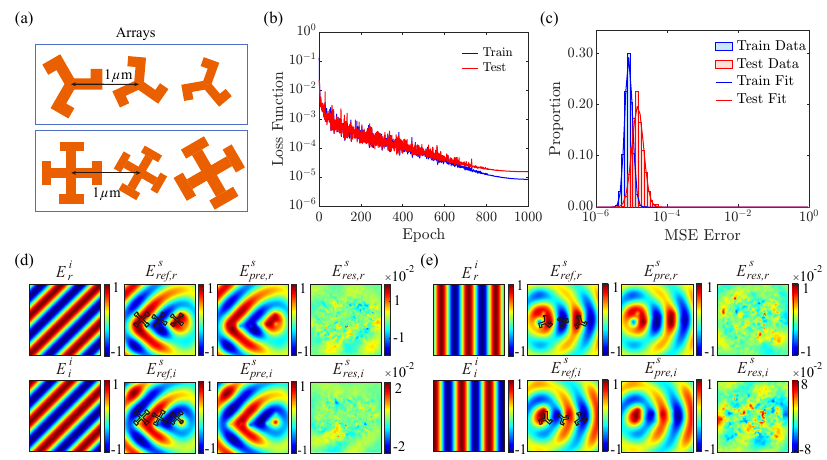}
    \caption{(a) Geometric structures of the scatterers. (b) Loss convergence curves. (c) MSE histograms with corresponding Gaussian fits. (d, e) Scattering fields for two representative multi-scatterer configurations. Columns (left to right): Incident field $E^i$, FEM reference $E_{ref}^s$, FEMONet prediction $E_{pre}^s$ and error $E_{res}^s$. Rows indicate the real and imaginary parts.}
    \label{fig5}
\end{figure}

We then examine multiple metallic scatterers to evaluate the numerical robustness of FEMONet under more strongly coupled scattering conditions. In this case, the active operator-parameter projections are the material and excitation subspaces, $\mathcal{P}_{\varepsilon_r}\times\mathcal{P}_{E^i}$. Compared with isolated scatterers, arrayed metallic structures introduce stronger near-field coupling and more complex interference patterns, making them a stringent test of the stability of the learned operator.

As shown in Fig.~\ref{fig5}(a), the structure is a $1\times3$ array assembled from two metallic scatterer units. Each scatterer undergoes random rotation and scaling around its center, with an adjacent spacing of $1~\mu\mathrm{m}$ in a $4~\mu\mathrm{m}$ square computational domain. A total of $3,456$ operator instances are generated, including $2,765$ training samples and $691$ testing samples. The convergence curves and MSE distributions in Fig.~\ref{fig5}(b,c) show stable training and consistent test performance. The representative fields in Fig.~\ref{fig5}(d,e) confirm that FEMONet accurately predicts the real and imaginary parts of the scattered fields even in the presence of multi-scatterer coupling.

\begin{figure}
    \centering
    \includegraphics[width=8.7cm]{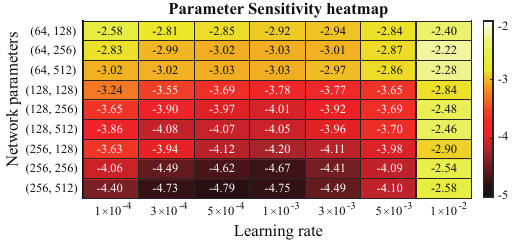}
    \caption{Heatmap analysis of $\log_{10}(\text{MSE})$ for different network hyperparameters.}
    \label{fig6}
\end{figure}

To further assess whether this performance depends on a carefully tuned network configuration, we perform a hyperparameter sensitivity analysis. As shown in Fig.~\ref{fig6}, nine architectural combinations are tested by varying the output dimension in $\{64,128,256\}$ and the hidden-channel number in $\{128,256,512\}$, together with seven initial learning rates from $10^{-4}$ to $10^{-2}$. The heatmap of $\log_{10}(\mathrm{MSE})$ reveals a broad and contiguous low-error region rather than a single isolated optimum. This indicates that FEMONet is not highly sensitive to a narrowly selected hyperparameter setting. The combination of stable convergence, accurate field reconstruction, and a broad low-MSE region demonstrates the numerical robustness of the FEM-constrained operator-learning framework.

\subsection{Sparse-sample learning of plasmonic responses}

\begin{figure}[t!]
    \centering
    \includegraphics[width=14cm]{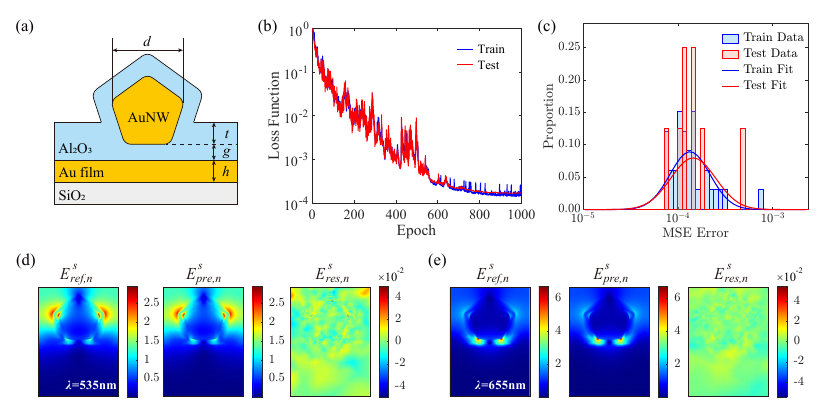}
    \caption{(a) Geometric structures of the scatterers. (b) Loss convergence curves. (c) MSE histograms with corresponding Gaussian fits. (d, e) Electric field norm distributions of the two resonant modes. Columns (left to right): FEM reference $E_{ref}^s$, FEMONet prediction $E_{pre}^s$ and error $E_{res}^s$.}
    \label{fig7}
\end{figure}

We next examine whether FEMONet can learn plasmonic responses from sparse-samples in the operator parameter space. The benchmark structure is adapted from a previously reported surface-plasmon-polariton device \cite{jiang2020temperature}, which has been validated by both numerical simulations and experimental measurements. This experimentally grounded configuration provides a more stringent test than idealized toy models. To ensure that these findings are not merely an isolated case, we provide another sparse-sample simulation of a practical device in SI Appendix\cite{le2024surface}.

The nanowire-on-mirror structure in Fig.~\ref{fig7}(a) consists of an Au nanowire with a pentagonal cross-section placed above a $200~\mathrm{nm}$ Au mirror, separated by a $5~\mathrm{nm}$ $\mathrm{Al}_2\mathrm{O}_3$ spacer and covered by a $5~\mathrm{nm}$ $\mathrm{Al}_2\mathrm{O}_3$ capping layer. The wavelength range is $500\sim800~\mathrm{nm}$, and the Au permittivity is described by the Drude model. Because the material response varies with wavelength, each sampled configuration corresponds to a coupled spectral-material state in $\mathcal{P}_{\lambda}\times\mathcal{P}_{\varepsilon_r}$. We use only $41$ wavelength-permittivity paired samples with a $5~\mathrm{nm}$ interval, including $33$ samples for training and $8$ for testing.

Despite the limited number of samples, the training and testing losses in Fig.~\ref{fig7}(b) stabilize at $1.5\times10^{-3}$ and $1.7\times10^{-3}$, respectively, and the MSE distributions in Fig.~\ref{fig7}(c) remain well concentrated. The moderate increase in absolute MSE is primarily attributable to two factors: first, the statistical constraints inherent to low-sample learning, and second, the strong localized plasmonic enhancement within the nanostructure. Thus, field-pattern fidelity provides an essential complementary measure for evaluating the learned operator in this resonant nanophotonic system.

The field maps in Fig.~\ref{fig7}(d,e) show that FEMONet accurately reproduces the electric-field norm at $\lambda=535~\mathrm{nm}$ and $\lambda=655~\mathrm{nm}$, corresponding to the transverse dipole mode and the lowest-order cavity plasmon mode, respectively. This case demonstrates the ability of FEMONet to accurately simulate the complex optical scattering characteristics of practical optical devices from sparse-samples of the coupled spectral-material operator parameter space.

\subsection{Three-dimensional metasurface operator learning}
\begin{figure}
    \centering
    \includegraphics[width=14cm]{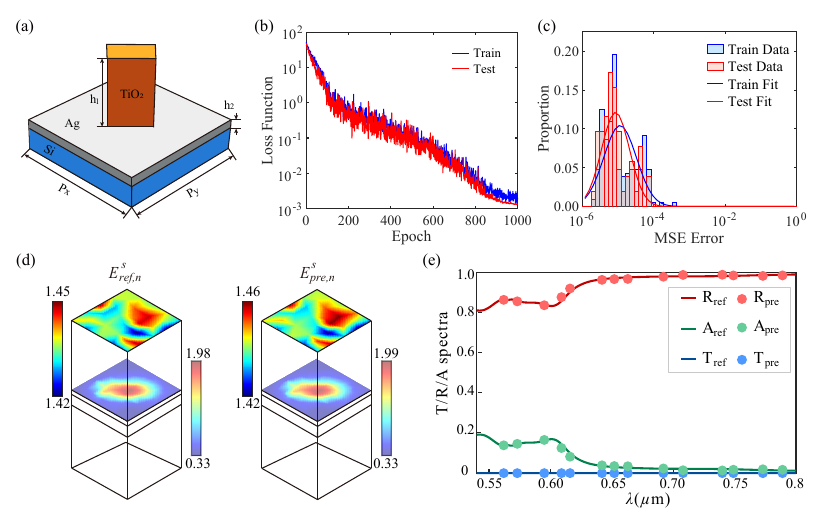}
    \caption{(a) Geometric structures of the scatterers. (b) Loss convergence curves. (c) MSE histograms with corresponding Gaussian fits.(d) Electric field maps at the incident and metal-dielectric surfaces. (left: FEM reference; right: predicted solution). (e) Transmittance, reflectance, and absorption spectra.}
    \label{fig8}
\end{figure}
Finally, we extend FEMONet to a three-dimensional metasurface unit cell\cite{fan2017visible} to demonstrate dimensional scalability. This case activates the spectral and material subspaces, $\mathcal{P}_{\lambda}\times\mathcal{P}_{\varepsilon_r}$, while replacing the two-dimensional basis representation with three-dimensional basis representation. It therefore tests whether the proposed framework can be generalized from two-dimensional scattering benchmarks to practical three-dimensional nanophotonic structures.

The unit cell in Fig.~\ref{fig8}(a) has a period of $P_x=P_y=250~\mathrm{nm}$ and consists of $\mathrm{TiO}_2$ nanofins on an Ag mirror. The $\mathrm{TiO}_2$ nanofin has $L=190~\mathrm{nm}$, $W=90~\mathrm{nm}$, and $h_1=600~\mathrm{nm}$, while the Ag mirror thickness is $h_2=150~\mathrm{nm}$. Material dispersion over $540-800~\mathrm{nm}$ is described using literature-based models for $\mathrm{TiO}_2$, Si, and Ag. The dataset contains $261$ wavelength configurations, split into $209$ training samples and $52$ testing samples.

The loss curves in Fig.~\ref{fig8}(b) show that both training and testing errors stabilize around $2\times10^{-3}$, and the MSE distributions in Fig.~\ref{fig8}(c) confirm generalization under the three-dimensional setting. Fig.~\ref{fig8}(d) compares the electric-field maps at the incident and metal-dielectric surfaces, showing close agreement between FEM reference fields and FEMONet predictions. More importantly, because FEMONet predicts the field distribution rather than directly fitting spectral coefficients, the predicted fields can be used with the Poynting theorem to compute reflection, transmission, and absorption spectra. As shown in Fig.~\ref{fig8}(e), the spectra derived from FEMONet agree with FEM reference results and satisfy the energy-conservation relation $R+A+T=1$. This demonstrates that FEMONet provides not only accurate field-level prediction, but also physically consistent optical-response estimation for practical 3D metasurfaces.
\subsection{Computational acceleration across optical scattering tasks}
\begin{table}[htbp]
\centering
\begin{threeparttable}
\caption{Online computational performance across validation cases}
\label{tab2}
\begin{tabular}{c c c c c c}
\toprule
\makecell[c]{Case} & \makecell[c]{Method}
& \makecell[c]{Average \\ Unknowns}
& \makecell[c]{MSE} 
& \makecell[c]{Solve Time \\ /Inference Time, \\Relative Percentage} 
& \makecell[c]{Training \\ Time} 
\\
\midrule
\multirow{3}{*}{\makecell[c]{1.Basic scatterers}} 
                   & FEM  & \multirow{3}{*}{383} & - &3.54$\times10^{-3}$s  &-  \\
                   & \makecell[c]{Enhanced VPINN}  & & $5.29\times10^{-4}$ &-&17h 28min  \\
                   & FEMONet & & $4.39\times10^{-5}$ &\makecell[c]{1.37$\times10^{-4}$s, 3.87$\%$ 
                   \\1.02 $\times10^{-4}$s, 2.88$\%$} & 2h 12min \\
\midrule
\multirow{2}{*}{\makecell[c]{2.Metal scatterers}} 
                   & FEM  &\multirow{2}{*}{532} & - &2.90$\times10^{-3}$s & -  \\
                   & FEMONet  & & $1.35\times10^{-3}$ &\makecell[c]{1.74$\times10^{-4}$s, 6.00$\%$ \\ 8.20$\times10^{-5}$s, 2.83$\%$} & 170h 42min  \\
\midrule
\multirow{2}{*}{\makecell[c]{3.Multiple metal scatterers}} 
                   & FEM  &\multirow{2}{*}{1451} & - &2.14$\times10^{-2}$s &-  \\
                   & FEMONet  & & $1.57\times10^{-5}$ &\makecell[c]{7.40$\times10^{-4}$s, 3.46$\%$ \\3.71$\times10^{-4}$s, 1.73$\%$ } &27h 23min\\
\midrule
\multirow{2}{*}{\makecell[c]{4.Plasmonic  nanostructure}} 
                   & FEM  &\multirow{2}{*}{3491} & - &1.61$\times10^{-2}$s & -\\
                   & FEMONet   & & $1.70\times10^{-4}$ & \makecell[c]{7.18$\times10^{-4}$s, 4.46$\%$ \\4.36$\times10^{-4}$s, 2.71$\%$} &  12min  \\
\midrule
\multirow{2}{*}{\makecell[c]{5.3D metasurface}} 
                   & FEM  &\multirow{2}{*}{20209} & - & 7.19s  & - \\
                   & FEMONet  & & $1.33\times10^{-5}$ & \makecell[c]{2.29$\times10^{-2}$s, 0.32$\%$ \\1.32$\times10^{-2}$s, 0.18$\%$} & 5h 38min\\ 
\bottomrule
\end{tabular}
\begin{tablenotes}
\footnotesize
\item The percentages denote the ratio of FEMONet inference time to FEM solve time. For each FEMONet row, the first and second inference time correspond to FP32 and FP16 inference, respectively. Cases 2 and 5 use four GPUs for parallel inference. FEM solve time excludes matrix assembly
\end{tablenotes}
\end{threeparttable}
\end{table}
After validating the accuracy and generality of FEMONet, we quantified its online computational efficiency against conventional FEM solvers across the five scattering problems. For the FEM baseline, the discrete linear system in Eq.~(\ref{Ab}) was solved using the BiCGSTAB solver in SciPy, with the stopping criterion adjusted to match the accuracy level achieved by FEMONet. The FEM timings were measured on an Intel Xeon Platinum 8352V CPU, whereas FEMONet inference was measured on NVIDIA GeForce RTX 4090 GPUs. This comparison is intended to reflect a deployment-oriented online simulation setting rather than a hardware-normalized theoretical comparison of numerical algorithms. To provide a conservative estimate of acceleration, the FEM timing includes only the linear-system solution time and excludes matrix assembly.

Table~\ref{tab2} summarizes the average number of unknown coefficients, prediction accuracy, online solve or inference time, and offline training time for all validation cases. FEMONet consistently reduces the online computational cost relative to FEM. Across the two-dimensional examples, the inference time is only $1.73\%\sim6.00\%$ of the FEM solve time, depending on the case and inference precision. The acceleration becomes more pronounced in the three-dimensional metasurface example, where FEMONet requires only $0.18\%\sim0.32\%$ of the FEM solve time. These results indicate that the benefit of operator learning increases as the electromagnetic problem size grows.

The offline training stage requires an upfront computational investment, especially for large operator-parameter spaces such as the metallic-scatterer case. However, once trained, FEMONet provides near-instantaneous predictions for new parameter instances within the prescribed operator parameter space. This offline-online trade-off is particularly advantageous for design, optimization, and repeated-query simulations, where a trained model can be reused many times. The results in Table~\ref{tab2} therefore show that FEMONet not only preserves accuracy across diverse scattering problems, but also transforms repeated full-wave simulations into a GPU-compatible inference task with substantial online acceleration.

\section{Discussion}
We have introduced FEMONet, an operator-parameter-space-augmented FEM-constrained operator-learning framework for optical scattering. The central idea is to reformulate full-wave simulation from repeatedly solving isolated forward problems into learning a family of parameterized scattering operators. In this framework, the operator parameter space is defined by the physical entities that specify the wave-equation problem, including wavelength, material distribution, incident field, and boundary conditions. Through the variational weak form and FEM discretization, these entities are encoded into the stiffness matrix and load vector, thereby linking the operator parameter space, coordinate space, and physical solution space.

Rather than simply enlarging the input space, the operator parameter space defines a physically sufficient domain on which neural-operator generalization is meaningful, interpretable, and bounded. In FEMONet, a physical quantity is treated as an active operator-parameter subspace only when it changes the scattering operator by entering the differential operator, source term, boundary condition, and consequently the assembled stiffness matrix $A$ and load vector $b$, ultimately altering the solution space. Quantities that remain fixed are instead incorporated into the invariant operator family rather than introduced as learnable inputs. This construction provides a physical criterion for selecting the input function spaces of operator learning and clarifies the scope of generalization: the trained model transfers among problem instances within the prescribed operator parameter space, rather than extrapolating indefinitely across arbitrary geometries, materials, or excitation.

This framework also clarifies why the network architecture and the physical constraint act coherently. MIONet provides a natural representation of the projected operator-parameter subspaces through multiple branch networks, while the trunk network represents the coordinate dependence. By predicting FEM expansion coefficients rather than unconstrained field values, FEMONet inherits the trial-space structure of the Galerkin finite-element formulation. The FEM residual then supplies a weak-form physical constraint that is consistent with the underlying numerical discretization. This distinguishes FEMONet from variational neural networks that use separately prescribed test spaces without explicitly enforcing Galerkin-compatible trial-space representations.

The numerical results demonstrate the effectiveness of this framework across dielectric scatterers, dispersive metallic scatterers, multiple-scattering configurations, experimentally grounded plasmonic nanostructures, and three-dimensional metasurfaces. The ablation study shows that the Galerkin-consistent FEM constraint improves both accuracy and training efficiency compared with the enhanced VPINN baseline. The multi-branch framework further allows the learned operator to cover coupled spectral, material, and excitation subspaces. Sparse-sample tests in plasmonic structures and field prediction in a three-dimensional metasurface show that the framework can preserve physically meaningful resonant and optical-response information.

The computational benchmark highlights the practical value of this offline-online framework. Although training requires an upfront cost, the trained model provides fast online inference for new parameter instances within the same operator parameter space. This property is particularly useful for repeated-query electromagnetic simulations, including parametric sweeps, device optimization, and inverse design. Because the output is a field distribution rather than only an end-to-end optical spectrum, physical quantities such as reflection, transmission, and absorption can be computed using physical laws, improving interpretability relative to purely data-driven response predictors.

Future work may extend this framework toward broader optical settings, including anisotropic, nonlinear, or multi-physics materials, adaptive sampling of high-error regions in the operator parameter space, and integration with inverse-design loops. On the computational side, lightweight inference architectures, reduced-precision deployment, pruning, and knowledge distillation could further reduce the online cost. These directions would expand FEMONet from a forward surrogate for optical scattering into a general physics-consistent operator-learning platform for accelerated nanophotonic modeling and design.

\section{Methods and Materials}
\subsection{Loss Function}
\begin{table}[t!]
\centering
\begin{threeparttable}
\caption{Various numerical methods based on different loss functions}
    \centering
    \begin{tabular}{cc}
    \toprule
    Methods& Loss functions \\
    \midrule
    PINNs &  $\frac{1}{N_u}\sum\limits_{i=1}^{N_u}|\mathcal{L}u_{\theta,i}-f|^2+\lambda_b\frac{1}{N_b}\sum\limits_{i=1}^{N_b}|u_{\theta,i}-h|^2$\\
    VPINNs & $\frac{1}{N_e}\sum\limits_{i=1}^{N_e}|((\mathcal{L}u_{\theta}-f),v_i)_\Omega|^2+\lambda_b\frac{1}{N_b}\sum\limits_{i=1}^{N_b}|((\mathcal{B}u_{\theta}-h),v_i)_{\partial\Omega}|^2$ \\
    FEMONet & $\frac{1}{N_s}\sum\limits_{s=1}^{N_s}\frac{1}{N_{p_s}}||M_{p_s}^{-1}\left(A_{p_s}u_{p_s,\theta}-b_{p_s}\right)||^2_2$ \\ 
    \bottomrule
\end{tabular}
\begin{tablenotes}
\footnotesize
\item$u_{\theta}$ denotes a continuous neural-network approximation of the field; $u_{p_s,\theta}$ denotes the FEM coefficient vector predicted by FEMONet for the sampled operator instance $p_s$. $v_i$ denotes a test function, $A_{p_s}$ and $b_{p_s}$ are the FEM stiffness matrix and load vector, and $M_{p_s}^{-1}$ is the preconditioner.
\end{tablenotes}
\label{table:Various}
\end{threeparttable}
\end{table}
Table~\ref{table:Various} compares the loss functions used in several representative physics-constrained numerical methods. Classical PINNs enforce the governing equation in its strong form at collocation points, together with a boundary penalty term. From the perspective of the weighted-residual method, this pointwise residual can be interpreted as using Dirac delta functions centered at the collocation points as test functions~\cite{raissi2019physics,VPINN,hpVPINN}. This mesh-free formulation is flexible, but it requires coordinate-based automatic differentiation of the neural-network output and can be difficult to optimize for oscillatory, high-frequency, or multi-scale PDEs~\cite{raissi2019physics,18WangUndersan,19wang2022and}.

VPINNs generalize this collocation-based residual by replacing the delta-type test functions with regular polynomial or element-wise test functions, leading to a weak residual in a Petrov-Galerkin sense. Through integration by parts, derivatives are transferred from the neural-network trial function to the test functions, thereby reducing the required differentiation order of the neural-network approximation~\cite{VPINN,hpVPINN}. However, weak-form enforcement alone does not automatically imply numerical stability. In standard VPINN-type formulations, the trial function is represented by a continuous neural network, whereas the test space is prescribed separately. Unless the resulting trial-test pairing satisfies a suitable stability condition, such as the inf-sup condition in Petrov-Galerkin methods, the weak residual may fail to control unresolved components of the neural-network approximation~\cite{ErnGuermond2004,BoffiBrezziFortin2013,Bathe2001}.

The potential limitation of a separately prescribed test space can be illustrated by a simple weak-residual example\cite{fang2021high}. Consider the trivial problem $u(x)=0$ on $[-1,1]$. If Legendre polynomials $P_j(x)$, $j=0,\ldots,K$, are selected as test functions, the VPINNs residual can be written as 
\begin{equation*} 
L(\theta) = \sum_{j=0}^{K} \left| \int_{-1}^{1} u_{\theta}(x)P_j(x)\,\mathrm{d}x \right|^2 . 
\end{equation*} 
Now suppose the neural-network approximation contains a component in the unresolved high-order subspace, for example $u_{\theta}(x) = \sum_{\ell=K+1}^{K+M} a_{\ell}P_{\ell}(x)$. By the orthogonality of Legendre polynomials, $L(\theta)=0$ even though $u_{\theta}(x)\neq0$. This example shows that VPINNs can be blind to components outside the span of the selected test functions. Meanwhile, the example highlights that weak-form neural residuals do not automatically provide Galerkin-type discrete stability unless the trial and test spaces are made compatible. 

Classical conforming FEM avoids this mismatch by constructing the trial and test spaces from the same finite-dimensional basis, leading to the discrete system $Au=b$. The derivatives of the basis functions, material parameters, source terms, and boundary contributions are assembled into the stiffness matrix and load vector. Thus, the weak-form constraint is enforced in a discrete space whose stability properties can be analyzed by finite-element theory. FEMONet transfers this Galerkin-consistent constraint into operator learning. Instead of directly approximating the continuous field by an unconstrained neural function, the network predicts the FEM expansion coefficients associated with a prescribed finite-element basis, i.e., 
$E_z^s(\boldsymbol{r};p_s) \approx \sum_{i=1}^{N_{p_s}} u_{p_s,\theta,i}N_i(\boldsymbol{r})$ and $u_{p_s,\theta,i} = \mathcal{G}_{s,\theta} (\lambda,\varepsilon_r,E^i) (\boldsymbol{r}_i)$. 

For each sampled operator instance $p_s$, the physical entities in the operator parameter space are encoded by the FEM matrix $A_{p_s}$ and load vector $b_{p_s}$. The training objective is therefore defined by the preconditioned FEM residual, 
\begin{equation} 
L(\theta) = \frac{1}{N_s} \sum_{s=1}^{N_s} \frac{1}{N_{p_s}} \left\| M_{p_s}^{-1} \left( A_{p_s}u_{p_s,\theta}-b_{p_s} \right) \right\|_2^2. 
\end{equation} 
This construction explains the advantage of FEMONet. First, the neural-network output belongs to the FEM trial space by construction, so the learned solution is constrained by the same Galerkin discretization used in classical FEM. Second, the physics loss avoids coordinate-based automatic differentiation because the spatial derivatives of the basis functions have already been evaluated during FEM assembly and stored in $A_{p_s}$. Third, the loss is evaluated over sampled operator instances $p_s\in\mathcal{P}$, which extends the FEM weak-form constraint from a single linear system to a family of parameterized scattering operators. Finally, the preconditioner $M_{p_s}^{-1}$ improves the condition number of the residual minimization for the sparse, complex-valued, and non-positive definite systems arising in optical scattering problems. Detailed derivations of the residual representation in complex-valued system and sparse form residual evaluation are provided in the SI Appendix.

\subsection{Network Architecture}
DeepONet was developed to learn nonlinear operators between function spaces. Given an input function $v:D\ni x\mapsto v(x)\in \mathbb{R}$ and an output function $u:D'\ni \boldsymbol{r}\mapsto u(\boldsymbol{r})\in \mathbb{R}$, the target operator can be written as
\begin{equation}
    \label{Gm}
    \mathcal{G}: \mathcal{V}\ni v \mapsto u \in \mathcal{U}, 
\end{equation}
where $\mathcal{V}$ and $\mathcal{U}$ denote the input and output function spaces, respectively. DeepONet represents $\mathcal{G}$ through a branch-trunk architecture: the branch network encodes the sampled values of $v$ at sensor points ${\xi_1,\xi_2,\ldots,\xi_m}$, whereas the trunk network encodes the query coordinate $\boldsymbol{r}$. The operator output is obtained by combining the branch and trunk features,
\begin{equation}
    \label{deeponet}
    \mathcal{G}(v)(r)=\sum_{k=1}^{q}b_k(v)t_k(r)+b_0, 
\end{equation}
where $b_0$ is a bias term, and $q$ is the latent dimension. This branch-trunk decomposition is well suited for optical scattering, because the branch network represents the physical inputs defining a scattering problem, while the trunk network represents the coordinate dependence of the field.

MIONet extends this construction to operators defined on the Cartesian product of multiple input-function spaces. For input functions $v_i:D_i\ni x\mapsto v_i(x)\in \mathbb{R}$, the multi-input operator is
\begin{equation}
    \label{MGm}
    \mathcal{G}:  \mathcal{V}_1\times...\times \mathcal{V}_n \mapsto \mathcal{U}, 
\end{equation}
where $\mathcal{V}_i$ denotes the function space of the $i^{th}$ input and $\mathcal{U}$ denotes the function space of the output. MIONet uses one branch network for each input function and a shared trunk network for the output coordinate. Its operator approximation is written as
\begin{equation}
    \label{MIONet}
    \mathcal{G}(v_1,...,v_n)(r)=\sum_{k=1}^{q}(\prod_{i=1}^{n} b_k^i(v_i))t_k(r)+b_0, 
\end{equation}
This multi-branch structure is naturally compatible with the operator parameter space in FEMONet, because different branches can encode different projected subspaces of $\mathcal{P}$, such as the spectral, material, and excitation subspaces.

Based on this formulation, FEMONet uses a MIONet architecture to learn the parameterized scattering operator over ($\mathcal{P}$), as shown in Fig.~\ref{fig2}. The key advantage of this multi-branch formulation is that each physical input can be encoded by a neural architecture matched to its characteristic. In optical scattering, the wavelength is a global spectral variable, the material and excitation fields are spatially distributed physical quantities, and the test point coordinates provide continuous spatial query. FEMONet therefore uses different encoders for these different physical quantities while retaining a unified MIONet fusion rule.

The wavelength branch encodes the normalized spectral input $\lambda$, or equivalently the free-space wavenumber $k_0$. In light scattering, $k_0$ enters the governing equation through phase factors $e^{ik_0\boldsymbol{r}}$ and equation term $k_0^2\varepsilon_r$, thereby controlling the oscillatory behavior and spectral resonances of the electromagnetic field. To represent this spectral dependence efficiently, the wavelength input is mapped into a Fourier-feature space\cite{cooley1965algorithm,tancik2020fourier} and then processed by a feed-forward neural network to produce a (q)-dimensional latent representation. The material and excitation branches share the same convolutional-residual architecture\cite{lecun1998gradient,he2016deep} and extract spatial features from the relative permittivity distribution $\varepsilon_r$ and the incident electric field $E^i$, respectively. In the two-dimensional cases, these inputs are represented as $128\times128$ image-like tensors, with real and imaginary components stored as separate channels when applicable; In three-dimensional cases, the material and excitation inputs are represented as $128\times128\times128$ volumetric tensors and processed by a three-dimensional convolutional neural network\cite{tran2015learning}. The corresponding branch input spaces can be regarded as functional representations of projected operator-parameter subspaces.

The trunk network encodes the coordinate descriptor associated with the $i^{th}$ FEM degree of freedom. For scalar-basis problems, $\boldsymbol{r}_i$ denotes the vertex coordinate associated with the $i^{th}$ nodal degree of freedom, e.g., $\boldsymbol{r}_i=(x_i,y_i)$ in two-dimensional cases. For vector-basis problems, $\boldsymbol{r}_i$ denotes the endpoint-coordinate descriptor of the edge associated with the $i^{th}$ vector-basis degree of freedom, for example $\boldsymbol{r}_i=(\boldsymbol{r}_i^{(1)},\boldsymbol{r}_i^{(2)})$. Unlike the material and excitation inputs, $\boldsymbol{r}_i$ is not an image-like field but a continuous spatial query. Therefore, FEMONet encodes it using a gated MLP architecutres\cite{dauphin2017language,liu2021pay,18WangUndersan}. The gated structure adaptively blends two coordinate-dependent feature embeddings through a learned gate, allowing the trunk network to represent both smooth global coordinate dependence and localized variations associated with the coordinate space $\Omega$.

As shown in Fig.~\ref{fig2}(d), the latent representations from the branch and trunk networks are fused through the MIONet inner product to predict the FEM expansion coefficient$u_{p,\theta,i}=\mathcal{G}_{s,\theta}(\lambda,\varepsilon_r,E^i)(\boldsymbol{r}_i)$. For complex-valued optical scattering problems, FEMONet adopts a two-channel output representation. The latent dimension is divided into two groups: the inner-product result from the first half of the latent channels gives the real part of the FEM coefficient, while that from the second half gives the imaginary part. Thus, the network prediction for the $i^{th}$ degree of freedom is written as $u_{p,\theta,i}=u^{(r)}_{p,\theta,i}+\mathrm{i}u^{(i)}_{p,\theta,i}$. This representation allows the real-valued MIONet architecture to predict the complex-valued coefficient vector used in the FEM residual. The predicted coefficient vector is then passed to the FEM-constrained residual loss defined in the preceding subsection. Therefore, this architecture provides the network realization of the proposed framework: branch networks encode the operator parameter space, the trunk network encodes the coordinate space, and the output corresponds to the FEM coefficient space.

In the branch networks, input functions can be represented by either aligned or unaligned sensors. We adopt the unaligned representation. For example, the sensor grid used to represent $\varepsilon_r$ and $E^i$ does not need to coincide with the FEM degree-of-freedom coordinates used as trunk inputs.

\subsection{Training Samples Generation}
The sample sets were generated by processing parameterized COMSOL Multiphysics project files with MATLAB scripts. For each sampled operator instance $p_s$, the scripts updated the corresponding geometry, material distribution, wavelength, incident field, and boundary settings according to the prescribed operator-parameter subspace. The FEM mesh was then generated in COMSOL, and the quantities required for FEMONet training were exported, including the sparse FEM stiffness matrix $A_{p_s}$, load vector $b_{p_s}$, degree-of-freedom coordinates, material distribution $\varepsilon_r$, incident field $E^i$, and wavelength $\lambda$. For each validation case, $80\%$ of the generated samples were randomly selected as the training set, and the remaining $20\%$ were used as the testing set.

The FEM reference solutions were not used as supervised labels during FEMONet training. They were computed independently only for validation and field-error evaluation. The prediction accuracy was evaluated by the mean squared error $\mathrm{MSE}=\frac{1}{N}\sum_{i=1}^{N}|E_{ref}-E_{pre}|^2$. The case-specific geometry ranges, wavelength ranges, material models, incident-field settings, and boundary conditions are summarized in the Results section and/or in the SI Appendix.

\subsection{Training and Evaluation Details}
FEMONet was implemented in PyTorch. Unless otherwise specified, the models were trained using the Adam optimizer for 1,000 epochs with a cosine-annealed learning-rate schedule. The initial learning rate was set to $1\times10^{-3}$ and gradually decays to $1\times10^{-5}$ during training. The wavelength input was normalized within the sampled spectral range, and the spatial coordinates were normalized according to the computational domain. The material distribution and incident field were represented as image-like tensors; complex-valued quantities were stored using separate real and imaginary channels.

For complex-valued optical scattering problems, the network output was represented by two channels corresponding to the real and imaginary parts of the FEM coefficient vector. The predicted coefficient vector was then used to evaluate the preconditioned FEM residual loss. Sparse FEM matrices were not converted into dense matrices during training; instead, the residual was evaluated through sparse matrix-vector multiplication. Training and evaluation were performed on a computing platform equipped with an Intel Xeon Platinum 8352V CPU and multiple NVIDIA GeForce RTX 4090 GPUs.

For the ablation study, we used an enhanced VPINN baseline in which the scattering boundary condition is incorporated into the weak-form residual rather than imposed through a separate boundary penalty. This baseline provides a stronger comparison than the original VPINN formulation while retaining the continuous-field neural-network approximation. Details of the enhanced VPINN formulation and the corresponding computational-efficiency analysis are provided in the SI Appendix.

For evaluating inference time, to ensure a reliable baseline, Eq.~(\ref{Ab}) was solved via the biconjugate gradient stabilized method BiCGSTAB in the SciPy library, with its stopping criterion set to match the MSE achieved by FEMONet. Before timing, the first test sample was run once as a warm-up and was excluded from the timing statistics. After this warm-up run, both BiCGSTAB and FEMONet were evaluated over the complete testing set, and the reported time corresponds to the average per-sample solver time for BiCGSTAB or the average per-sample inference time for FEMONet. Various precisions (e.g., FP32, FP64, FP16, and BF16) can be deployed during inference; here, FP16 and FP32 were considered. Since the minimum positive representable value of FP16 is approximately $5.96 \times 10^{-8}$, it provides sufficient dynamic range for our numerical domain without compromising prediction accuracy. Crucially, while standard FEM includes both matrix assembly and system solution times, only the solution time is accounted for here to provide a conservative lower bound for the acceleration of FEMONet.

\section*{Acknowledgments}
The source code, sample-generation scripts, training and evaluation scripts, and datasets supporting the findings of this study have been deposited in GitHub and are publicly available at: \href{https://github.com/HUST-CPO/FEMONet-Robust-and-universal-Operator-learning-for-optical-scattering-problem-via-MIONet}{https://github.com/HUST-CPO/FEMONet-Robust-and-universal-Operator-learning-for-optical-scattering-problem-via-MIONet}\cite{GitHub}.

\appendix
\setcounter{equation}{0}
\setcounter{table}{0}
\setcounter{figure}{0}
\setcounter{section}{0}

\renewcommand{\thesection}{S\arabic{section}}
\renewcommand{\theequation}{S\arabic{equation}}
\renewcommand{\thetable}{S\arabic{table}}
\renewcommand{\thefigure}{S\arabic{figure}}

\section{appendix}
\section{Variational principle and finite element method}
Taking the two-dimensional optical scattering problem as an example, under transverse magnetic (TM) polarization, the electric field component $E_z$ satisfies the vector wave equation,
\begin{equation}
\label{eqn1}\nabla\times\overline{\mu_r}^{-1}\nabla\times\hat{z}E_z(\boldsymbol{r})-k_0^2\overline{\varepsilon}_r\hat{z}E_z(\boldsymbol{r})=0 \quad on \quad \Omega, 
\end{equation}
where $\nabla\times$ is the curl operator, $\overline{\mu_r}^{-1}$ is the relative permeability, $\overline{\varepsilon}_r$ is the relative permittivity, $k_0$ is the wavenumber in a vacuum, $\boldsymbol{r}=[x,y]^T$ is the spatial coordinates, $\Omega$ is the computational domain. The total field $E_z$ is the sum of incident field $E_z^i$ and the scattered field $E_z^s$, i.e, $E_z=E_z^i+E_z^s$. Thus, Eq.~(\ref{eqn1}) can be written as:
\begin{equation}
    \label{eqn2}
    \nabla\times\overline{\mu_r}^{-1}\nabla\times \hat{z}E_z^s(\boldsymbol{r})-k_0^2\overline{\varepsilon}_r\hat{z}E_z^s(\boldsymbol{r})=
    -\nabla\times\overline{\mu_r}^{-1}\nabla\times \hat{z}E_z^i(\boldsymbol{r})+k_0^2\overline{\varepsilon}_r\hat{z}E_z^i(\boldsymbol{r}) \quad on \quad \Omega.
\end{equation}
To simulate the wave propagation without reflection, the boundaries of the computational domain satisfy the scattering boundary condition
\begin{equation}
\label{bc}
\boldsymbol{n}\times\nabla\times\hat{z}E_z(\boldsymbol{r})-ik_0\boldsymbol{n}\times\hat{z}E_z^s(\boldsymbol{r})\times\boldsymbol{n}=0 ~ on ~\delta\Omega,
\end{equation}
where $\boldsymbol{n}$ is the outward unit normal vector of the boundary, $\delta\Omega$ is the outer boundary of $\Omega$. Eq.~(\ref{eqn2}) and  (\ref{bc}) corresponding to the boundary-value problems represent in Eq.~(\ref{boundary-value_porblems}).

Variational principle can be used to solve this differential equation in the domain $\Omega$:

\begin{equation}
    \mathcal{L}\phi=f, 
\end{equation}
where $\mathcal{L}=\nabla\times\left(\overline{\mu}_r^{-1}\nabla\times\right)-k_0^2\overline{\varepsilon}_r$, $\phi=E_z^s(\boldsymbol{r})$, and $f=-\nabla\times\overline{\mu}_r^{-1}\nabla\times \hat{\boldsymbol{z}}E_z^i(\boldsymbol{r})+k_0^2\overline{\varepsilon}_r\hat{\boldsymbol{z}}E_z^i(\boldsymbol{r})$, together with boundary conditions on the boundary $\partial \Omega$ that enclose the domain:
\begin{equation}
    \mathcal{B}\phi(\boldsymbol{r})=h(\boldsymbol{r}), \quad \boldsymbol{r}\in\partial \Omega.
\end{equation}
If the operator $\mathcal{L}$ is self-adjoint——that is,
\begin{equation}
    \label{eqns1-1}
\langle\mathcal{L}\phi,\psi\rangle=\langle\phi,\mathcal{L}\psi\rangle,
\end{equation}
and positive definite--that is,
\begin{equation}
    \label{eqns1-2}
    \langle\mathcal{L}\phi,\phi\rangle
\begin{cases}
>0 & \quad\phi\neq0 \\
=0 & \quad\phi=0, 
\end{cases}
\end{equation}
then its solution can be obtained by minimizing the functional given by
\begin{equation}
    \label{eqns1-3}
    F(\phi)=\frac{1}{2}\langle\mathcal{L}\phi,\phi\rangle-\frac{1}{2}\langle\phi,f\rangle-\frac{1}{2}\langle f,\phi\rangle.
\end{equation}
In these equations, $\psi$ denotes an arbitrary function satisfying the same boundary conditions as does $\phi$. The angular brackets denote the inner product in Hilbert space:
\begin{equation}
    \label{eqns1-4}
    \langle\phi,\psi\rangle=\int_{\Omega}\phi\psi^*\mathrm{d}\Omega,
\end{equation}
where the asterisk denotes complex conjugate operation. Under this definition of the inner product, the conditions for the operator $\mathcal{L}$ defined by Eq.~(\ref{eqn3}) to be self-adjoint are (a) $\overline{\varepsilon}_r,\overline{\mu_r}$ are real numbers or functions and (b) the boundary conditions are homogeneous. According to the definition of the inner product in Hilbert space,
\begin{equation}
    \label{eqns1-5}
    \langle\mathcal{L}\boldsymbol{E},\boldsymbol{E}\rangle=\int_{V}[\frac{1}{\overline{\mu_r}}(\nabla\times\boldsymbol{E})\cdot(\nabla\times\boldsymbol{E})^*-k_0^2\overline{\varepsilon}_r\boldsymbol{E}\cdot\boldsymbol{E}^*]\mathrm{d}V, \hat{n}\times\boldsymbol{E}=0, \:at\: \partial V
\end{equation}
which not be positive, and therefore the operator $\mathcal{L}$ be non-positive definite. However, notice that the condition that limits self-adjoint operators to real operators is a direct consequence of the definition of the inner product. If we define the inner product as symmetry inner product,
\begin{equation}
    \label{eqns1-6}
    \langle\phi,\psi\rangle=\int_{\Omega}\phi\psi\mathrm{d}\Omega,
\end{equation}
then the limitation is lifted immediately. The solution of Eq.~(\ref{eqn3}) can be obtained by finding the stationary point of the functional $F(\phi)$, which is equivalent to solving
\begin{equation}
    \label{eqns1-7}
    \partial F= \frac{1}{2}\langle\mathcal{L}\delta\phi,\phi\rangle+\frac{1}{2}\langle\mathcal{L}\phi,\delta\phi\rangle-\frac{1}{2}\langle\delta\phi,f\rangle-\frac{1}{2}\langle f,\delta\phi\rangle=0.
\end{equation}

This distinction is mathematically fundamental for optical scattering; it explicitly precludes the application of traditional physics-informed neural networks (such as the Deep Ritz method) that rely on minimizing energy functionals\cite{ew2018deep,rezaei2025finite}. Since solving this infinite-dimensional variational problem directly is generally intractable, the Ritz-Galerkin method introduces a trial function $\tilde{\phi}$ in a finite-dimensional subspace to approximate $\phi$, which can be approximated by the expansion 
\begin{equation}
    \tilde{\phi}=\sum_{i=1}^Nc_iv_i, 
\end{equation}
where $v_i$ are the chosen expansion functions and $c_i$ are constant coefficients to be determined. In FEM, this finite-dimensional subspace is constructed by discretizing the domain $\Omega$ into non-overlapping elements and defining local shape functions over these elements. To find the stationary points of $F(\tilde{\phi})$, we force its partial derivatives with respect to $c_i$ to vanish. This yields a set of linear algebraic equations
\begin{equation}
    \label{partial}
    \frac{\partial F}{\partial c_i}=\frac{1}{2}\sum_{j=1}^{N}c_j\int_\Omega (v_i\mathcal{L}v_j+v_j\mathcal{L}v_i)d\Omega-\int_\Omega v_ifd\Omega \quad i=1,2,...,N,
\end{equation}
where the test function $v_j$ is introduced to enforce the stationary condition, and this equation can be written as a matrix form:
\begin{equation}
\label{Ab2}
    Au=b. 
\end{equation}
For the problem described by Eq.~(\ref{eqn2}) and Eq.~(\ref{bc}), the electric field is expanded using discrete basis functions $E_s^z=\Sigma^N_{i=1}u_iN_i$, where $u_i$ is the DOF to be solved, $N_i$ is the basis function, and $N$ is the total number of DOFs. The elements in $A$ given by
\begin{equation}
\label{eqnaij}
      A_{ij}=\frac{1}{2}\int_\Omega (v_i\mathcal{L}v_j+v_j\mathcal{L}v_i)d\Omega
      =\int_\Omega[\nabla\times\hat{z}N_i \cdot \overline{\mu_r}^{-1}\nabla\times \hat{z}N_j-k_0^2\overline{\varepsilon}_r\hat{z}N_i\cdot\hat{z}N_j]dS+\int_{\delta\Omega}[\hat{z}N_i\cdot jk_0\boldsymbol{n}\times\hat{z}N_j\times\boldsymbol{n} ]dl,
\end{equation}
and the elements in $b$ given by
\begin{equation}
    \label{bij}
        b_i=\int_\Omega v_ifd\Omega
        =\int_{\delta\Omega}[\hat{z}N_i\cdot\boldsymbol{n}\times\nabla\times\hat{z}E_z^i]dl
        -\int_\Omega[\nabla\times\hat{z}N_i\cdot\overline{\mu_r}^{-1}\nabla\times\hat{z}E_z^i-k_0^2\overline{\varepsilon}_r\hat{z}N_i\cdot\hat{z}E_z^i]dS.
\end{equation}
FEM obtains constant coefficient $u_i$ by solving the sparser linear system in Eq.(~\ref{Ab2}), and then obtains the solution in $\Omega$ by combining chosen expansion function $v_i$.

\section{Deep neural networks for solving linear system}
Since the linear system Eq.~(\ref{Ab2}) is derived from a physical problem having a smooth unknown function $v$, it is expected that the grid mapping $\chi$ defined by $\chi (x_j,y_j) = u_j,j=1,...,N$ 
is spatially smooth (it means the data $\{(x_j,y_j),u_j\}$  can be fit by a function with
few high-frequency components). Thanks to the good approximation property for
high-dimensional functions (see, e.g., [30, 48, 49, 9, 18, 45, 36, 46, 47]), NNs can be
employed to serve as the functioning of $\chi$. Specifically, we introduce an NN $\mathcal{N}_\theta(x,y)$ to approximate the mapping $\chi$, such that $\mathcal{N}_\theta(x,y) \approx u_j,j=1,...,N$. 
By this setting, and let
\begin{equation}
    \label{eqns2-1}
    \mathcal{N}_\theta:=[\mathcal{N}_\theta(x_1,y_1),\mathcal{N}_\theta(x_2,y_2),...,\mathcal{N}_\theta(x_N,y_N)]^T,
\end{equation}
the linear system Eq.~(\ref{Ab2}) can be formulated into
\begin{equation}
    \label{eqns2-2}
    A\mathcal{N}_\theta=b.
\end{equation}
Usually, the system Eq.~(\ref{eqns2-2}) does not have an exact solution $\theta$. So we will find the least square solution of Eq.~(\ref{eqns2-2}) through the following optimization framework:
\begin{equation}
    \label{eqns2-3}
    \min_{\theta}L(\theta)=\frac{1}{N}||A\mathcal{N}_\theta-b||_2^2.
\end{equation}
Since the definition of the symmetric inner product in some physical models may lead to a complex-valued linear system, we consider the case that $A \in \mathbb{C}^{N\times N}$ and $b\in \mathbb{C}^{N}$. Thus the residual vector $r(\theta):=A\mathcal{N}_\theta-b$ is complex.
For a complex vector $z\in\mathbb{C}^N$, we use the standard Hermitian norm
\begin{equation}
    \label{eqns2-4}
    \|z\|_2^2 := z^H z=\sum_{j=1}^N |z_j|^2,
\end{equation}
where $(\cdot)^H$ denotes the conjugate transpose. In order to measure the error in a fair way independent of the vector size, we adopt the scaled $\ell_2$-norm defined by
\begin{equation}
    \label{eqns2-5}
    \|z\|_{\ell_2}^2:=\frac{1}{N}\|z\|_2^2.
\end{equation}
Therefore, the least-squares objective Eq.~(\ref{eqns2-3}) can be equivalently written as
\begin{equation}
    \label{eqns2-6}
    \min_{\theta} L(\theta)
    =\frac{1}{N}\|A\mathcal{N}_\theta-b\|_2^2
    =\|A\mathcal{N}_\theta-b\|_{\ell_2}^2.
\end{equation}
Since common neural network architectures output real numbers, we employ a two-channel representation to approximate the complex solution, i.e.,
\begin{equation}
    \label{eqns2-7}
    \mathcal{N}_\theta=\mathcal{N}_{\theta,r}+\mathrm{i}\mathcal{N}_{\theta,i}, \in\mathbb{C}^N
\end{equation}
where $\mathcal{N}_{\theta,r}$ and $\mathcal{N}_{\theta,i}$ are two output channels of one network.
Accordingly, we define the stacked vectors
\begin{equation}
    \label{eqns2-8}
    \mathcal{N}_{\theta,r}:=[\mathcal{N}_{\theta,r}(x_1,y_1),...,\mathcal{N}_{\theta,r}(x_N,y_N)]^T,\qquad
    \mathcal{N}_{\theta,i}:=[\mathcal{N}_{\theta,i}(x_1,y_1),...,\mathcal{N}_{\theta,i}(x_N,y_N)]^T.
\end{equation}

Let $A=A_{r}+\mathrm{i}A_{i}$ and $b=b_{r}+\mathrm{i}b_{i}$ with $A_{r},A_{i}\in\mathbb{R}^{N\times N}$ and $b_{r},b_{i}\in\mathbb{R}^N$.
Then the residual $r(\theta)=A\mathcal{N}_\theta-b$ can be decomposed into real and imaginary parts:
\begin{equation}
    \label{eqns2-9}
    r_{r}(\theta)=A_{r}\mathcal{N}_{\theta,r}-A_{i}\mathcal{N}_{\theta,i}-b_{r},\quad
    r_{i}(\theta)=A_{i}\mathcal{N}_{\theta,r}+A_{r}\mathcal{N}_{\theta,i}-b_{i}.
\end{equation}
Therefore, the loss function Eq.~(\ref{eqns2-6}) for the complex system can be equivalently expressed in purely real form:
\begin{equation}
    \label{eqns2-10}
    L(\theta)=\|r_{r}(\theta)\|_{\ell_2}^2+\|r_{i}(\theta)\|_{\ell_2}^2.
\end{equation}
From \eqref{eqns2-10}, the partial derivatives with respect to the real and imaginary outputs are
\begin{equation}
    \label{eqns2-11}
    \frac{\partial L}{\partial \mathcal{N}_{\theta,r}}
    =\frac{2}{N}\Big(A_{r}^T r_{r}+A_{i}^T r_{i}\Big),\qquad
    \frac{\partial L}{\partial \mathcal{N}_{\theta,i}}
    =\frac{2}{N}\Big(-A_{i}^T r_{r}+A_{r}^T r_{i}\Big).
\end{equation}
Thus, by the chain rule, denote the Jacobian matrices of the real and imaginary
network outputs with respect to the trainable parameters by
\begin{equation}
    \label{eqns2-12}
    J_r(\theta):=\frac{\partial \mathcal{N}_{\theta,r}}{\partial \theta}
    \in \mathbb{R}^{N\times q},
    \qquad
    J_i(\theta):=\frac{\partial \mathcal{N}_{\theta,i}}{\partial \theta}
    \in \mathbb{R}^{N\times q},
\end{equation}
where $q$ is the number of trainable parameters. Then
\begin{equation}
    \label{eqns2-13}
    \nabla_{\theta} L(\theta)
    =
    J_r(\theta)^T
    \frac{\partial L}{\partial \mathcal{N}_{\theta,r}}
    +
    J_i(\theta)^T
    \frac{\partial L}{\partial \mathcal{N}_{\theta,i}} .
\end{equation}
Substituting Eq.~(\ref{eqns2-11}) into Eq.~(\ref{eqns2-13}), we obtain
\begin{equation}
    \label{eqns2-14}
    \begin{aligned}
    \nabla_{\theta} L(\theta)
    =
    \frac{2}{N} J_r(\theta)^T
    \left(A_r^T r_r + A_i^T r_i\right)
    +
    \frac{2}{N} J_i(\theta)^T
    \left(-A_i^T r_r + A_r^T r_i\right).
    \end{aligned}
\end{equation}
Equivalently, define the augmented real-valued quantities
\begin{equation}
    \label{eqns2-15}
    \widetilde{\mathcal{N}}_\theta
    :=
    \begin{bmatrix}
        \mathcal{N}_{\theta,r}\\
        \mathcal{N}_{\theta,i}
    \end{bmatrix},
    \qquad
    \widetilde{b}
    :=
    \begin{bmatrix}
        b_r\\
        b_i
    \end{bmatrix},
    \qquad
    \widetilde{A}
    :=
    \begin{bmatrix}
        A_r & -A_i\\
        A_i & A_r
    \end{bmatrix}.
\end{equation}
Then the augmented residual is given by
\begin{equation}
    \label{eqns2-16}
    \widetilde{r}(\theta)
    =
    \widetilde{A}\widetilde{\mathcal{N}}_\theta-\widetilde{b}
    =
    \begin{bmatrix}
        r_r(\theta)\\
        r_i(\theta)
    \end{bmatrix}.
\end{equation}
Therefore, the loss function can be written as
\begin{equation}
    \label{eqns2-17}
    L(\theta)
    =
    \frac{1}{N}
    \left\|
    \widetilde{A}\widetilde{\mathcal{N}}_\theta-\widetilde{b}
    \right\|_2^2 .
\end{equation}
Furthermore, define the augmented Jacobian matrix as
\begin{equation}
    \label{eqns2-18}
    \widetilde{J}(\theta)
    :=
    \frac{\partial \widetilde{\mathcal{N}}_\theta}{\partial \theta}
    =
    \begin{bmatrix}
        \dfrac{\partial \mathcal{N}_{\theta,r}}{\partial \theta}\\[2mm]
        \dfrac{\partial \mathcal{N}_{\theta,i}}{\partial \theta}
    \end{bmatrix}
    \in \mathbb{R}^{2N\times q}.
\end{equation}
Hence, the gradient of the loss function can be compactly expressed as
\begin{equation}
    \label{eqns2-19}
    \nabla_{\theta} L(\theta)
    =
    \frac{2}{N}
    \widetilde{J}(\theta)^T
    \widetilde{A}^T
    \left(
    \widetilde{A}\widetilde{\mathcal{N}}_\theta-\widetilde{b}
    \right),
\end{equation}
which is a purely real expression and can be directly computed via back-propagation.

\section{Enhanced VPINN and computational efficiency of FEMONet}

Before comparing the computational efficiency of FEMONet, we first clarify the enhanced VPINN baseline used in this work. In the original VPINN formulation, a neural network directly approximates the scattered field,
\begin{equation}
E_z^s(\mathbf{r})\approx\tilde{u}_{\theta}(\mathbf{r}),
\end{equation}
where the input is the spatial coordinate $\mathbf{r}=(x,y)$, and the outputs are the real and imaginary parts of the scattered field. Given a set of admissible test functions $\{v_k\}_{k=1}^{K}$, the governing equation is multiplied by each test function and integrated over the computational domain, yielding the variational residual
\begin{equation}
\mathcal{R}_k=
\left(\mathcal{L}\tilde{u},v_k\right)_{\Omega}-\left(f,v_k\right)_{\Omega},\qquad k=1,2,\ldots,K,
\end{equation}
where $\tilde{u}$ denotes the neural-network approximation of the scattered field.

The classical VPINN loss consists of the variational residual loss and a separate boundary penalty term,
\begin{equation}
\mathcal{L}^{\mathrm{VPINN}}=\mathcal{L}_{R}^{\mathrm{VPINN}}+\mathcal{L}_{u},
\end{equation}
where
\begin{equation}
\begin{aligned}
\mathcal{L}_{R}^{\mathrm{VPINN}}&=\frac{1}{K}\sum_{k=1}^{K}\left|\mathcal{R}_k\right|^2\\
&=\frac{1}{K}\sum_{k=1}^{K}\left|\left(\mathcal{L}\tilde{u},v_k\right)_{\Omega}-\left(f,v_k\right)_{\Omega}\right|^2,
\end{aligned}
\end{equation}
and
\begin{equation}
\mathcal{L}_{u}=\alpha\frac{1}{N_u}\sum_{i=1}^{N_u}\left|r^{b}(\mathbf{x}_{u_i})\right|^2 .
\end{equation}
Here, $r^{b}(\mathbf{x}_{u_i})$ denotes the boundary residual evaluated at the boundary points $\{\mathbf{x}_{u_i}\}_{i=1}^{N_u}$, and $\alpha$ is the boundary-loss weight. Although the weak-form residual reduces the differential order required by the neural-network output, the boundary condition is still imposed separately. This may introduce a weight-balancing issue between the domain residual and the boundary residual and requires additional boundary evaluations during training.

To construct a stronger and fairer VPINN-type baseline, we adopt a enhanced VPINN formulation in which the scattering boundary condition is directly incorporated into the weak form. For Eq.~(\ref{eqn2}) and Eq.~(\ref{bc}), multiplying the governing equation by a test function $v_k$ and applying integration by parts gives
\begin{equation}
\int_{\Omega}\nabla \tilde{u}\cdot\nabla v_k\,\mathrm{d}\Omega-\int_{\partial\Omega}v_k\frac{\partial \tilde{u}}{\partial n}\,\mathrm{d}\Gamma-\int_{\Omega}k_0^2\varepsilon_r v_k\tilde{u}\,\mathrm{d}\Omega=\int_{\Omega}k_0^2(\varepsilon_r-1)v_k u^{\mathrm{inc}}\,\mathrm{d}\Omega .
\end{equation}
As a side remark, the curl--curl term in Eq.~(\ref{eqn2}) can be rewritten using the vector identity
\begin{equation}
\nabla\times(\nabla\times\mathbf{E})=\nabla(\nabla\cdot\mathbf{E})-\nabla^2\mathbf{E}.
\end{equation}
After integration by parts, the VPINN residual can therefore be expressed in a gradient-based form. In the present scalar formulation, this representation is mathematically equivalent to the curl-based form and is more convenient for component-wise automatic differentiation in neural-network implementation.

For the first-order scattering boundary condition,
\begin{equation}
\left(\frac{\partial}{\partial n}+\mathrm{i}k_0\right)u=0 \quad \text{on } \partial\Omega,
\end{equation}
we have
\begin{equation}
\frac{\partial u}{\partial n}=-\mathrm{i}k_0u .
\end{equation}
Substituting this relation into the boundary term yields
\begin{equation}
\int_{\Omega}\nabla \tilde{u}\cdot\nabla v_k\,\mathrm{d}\Omega+\int_{\partial\Omega}\mathrm{i}k_0v_k\tilde{u}\,\mathrm{d}\Gamma-\int_{\Omega}k_0^2\varepsilon_r v_k\tilde{u}\,\mathrm{d}\Omega=\int_{\Omega}k_0^2(\varepsilon_r-1)v_k u^{\mathrm{inc}}\,\mathrm{d}\Omega .
\end{equation}
Thus, for each test function $v_k$, the enhanced VPINN residual is defined as
\begin{equation}
\mathcal{R}_k^{\mathrm{E\text{-}VPINN}}=L_k-R_k,\end{equation}
where
\begin{equation}
L_k=\int_{\Omega}\nabla \tilde{u}\cdot\nabla v_k\,\mathrm{d}\Omega+\int_{\partial\Omega}\mathrm{i}k_0v_k\tilde{u}\,\mathrm{d}\Gamma-\int_{\Omega}k_0^2\varepsilon_r v_k\tilde{u}\,\mathrm{d}\Omega,
\end{equation}
and
\begin{equation}
R_k=\int_{\Omega}k_0^2(\varepsilon_r-1)v_k u^{\mathrm{inc}}\,\mathrm{d}\Omega .
\end{equation}
The enhanced VPINN loss is then written as
\begin{equation}
\begin{aligned}
\mathcal{L}_{\mathrm{E\text{-}VPINN}}=\frac{1}{K}\sum_{k=1}^{K}\left|\mathcal{R}_k^{\mathrm{E\text{-}VPINN}}\right|^2=\frac{1}{K}\sum_{k=1}^{K}\left|L_k-R_k\right|^2 .
\end{aligned}
\end{equation}

In this formulation, the scattering boundary condition is consistently embedded into the weak-form residual rather than treated as a separate penalty term, resulting in a more compact and physically consistent VPINN baseline. Under the same FNN architecture and single-GPU training environment, enhanced VPINN was evaluated on an isolated problem instance, consistent with the single-instance setting of the original VPINN formulation. Compared with the original VPINN, enhanced VPINN reduces the mean squared error (MSE) from $3.85\times10^{-5}$ to $3.18\times10^{-5}$ and shortens the training time from $260.13~\mathrm{s}$ to $180.94~\mathrm{s}$, corresponding to reductions of $17.4\%$ and $30.4\%$, respectively. Since this modification improves the loss construction without altering the continuous-field neural-network approximation, enhanced VPINN is adopted as a stronger baseline for evaluating the computational efficiency of FEMONet.

Since $A$ is fixed for a given geometry after FEM assembly, the training process only requires evaluating the sparse matrix-vector product $A\mathcal{N}_\theta$. For a sparse matrix A, let its nonzero entries be represented by
\begin{equation}
    \label{eqns3-1}
    \mathcal{S_A}=\{(i_m,j_m,a_m\}_{m=1}^{nnz(A)},
\end{equation}
where $i_m$ is the row index, $j_m$ is the column index, and
\begin{equation}
    \label{eqns3-2}
    a_m=A_{i_mj_m}.
\end{equation}
The matrix-vector product
\begin{equation}
    \label{eqns3-3}
    y=A\mathcal{N_\theta}
\end{equation}
has the component form
\begin{equation}
    \label{eqns3-4}
    y_i=\sum_{j=1}^NA_{ij}\mathcal{N}_{\theta,j}.
\end{equation}
Using only the nonzero entries of $A$, the same operation can be written as
\begin{equation}
    \label{eqns3-5}
    y_i=\sum_{m:i_m=i}a_{m}\mathcal{N}_{\theta,j}.
\end{equation}
Therefore, sparse matrix multiplication can be achieved through efficient vector multiplication. The residual is then obtained as
\begin{equation}
    \label{eqns3-6}
    r(\theta) = y-b = A\mathcal{N}_\theta -b.
\end{equation}
For mini-batch training, suppose a batch contains B samples. The FEM system of the s-th sample is
\begin{equation}
    \label{eqns3-7}
    A^{(s)}\mathcal{N}_\theta^{(s)}=b^{(s)}, s=1,2,...,B,
\end{equation}
The batch-level network output and source vector are obtained by concatenation:
\begin{equation}
\label{eqns3-8}
    \mathcal{N}^{\mathrm{bat}}_\theta
    =
    \begin{bmatrix}
        \mathcal{N}^{(1)}_\theta\\
        \mathcal{N}^{(2)}_\theta\\
        \vdots\\
        \mathcal{N}^{(B)}_\theta
    \end{bmatrix},
    \qquad
    b^{\mathrm{bat}}
    =
    \begin{bmatrix}
        b^{(1)}\\
        b^{(2)}\\
        \vdots\\
        b^{(B)}
    \end{bmatrix}.
\end{equation}
The corresponding batch matrix can be regarded as an implicit block-diagonal sparse matrix:
\begin{equation}
\label{eqns3-9}
    A^{\mathrm{bat}}
    =
    \begin{bmatrix}
        A^{(1)} & 0 & \cdots & 0\\
        0 & A^{(2)} & \cdots & 0\\
        \vdots & \vdots & \ddots & \vdots\\
        0 & 0 & \cdots & A^{(B)}
    \end{bmatrix}.
\end{equation}
Thus, the batch residual is
\begin{equation}
\label{eqns3-10}
    r^{\mathrm{bat}}(\theta)
    =
    A^{\mathrm{bat}}
    \mathcal{N}^{\mathrm{bat}}_\theta
    -
    b^{\mathrm{bat}} .
\end{equation}
The batch loss is
\begin{equation}
\label{eqns3-11}
    L^{\mathrm{bat}}(\theta)
    =
    \frac{1}{N_{\mathrm{bat}}}
    \left\|
    A^{\mathrm{bat}}
    \mathcal{N}^{\mathrm{bat}}_\theta
    -
    b^{\mathrm{bat}}
    \right\|_2^2 ,
\end{equation}
where
\begin{equation}
\label{eqns3-12}
    N_{\mathrm{bat}}
    =
    \sum_{s=1}^{B}N_s .
\end{equation}
In practice, $A^{\mathrm{bat}}$ is not explicitly constructed. Instead, the local sparse indices of each sample are shifted by a cumulative offset. For the $s$-th sample, define
\begin{equation}
\label{eqns3-13}
    o_s=\sum_{t=1}^{s-1}N_t .
\end{equation}
If a nonzero entry of $A^{(s)}$ is
\begin{equation}
\label{eqns3-14}
    A^{(s)}_{pq}=a^{(s)}_{pq},
\end{equation}
where $p$ is the local row index and $q$ is the local column index, then its global row and column indices in the batch system are
\begin{equation}
\label{eqns3-15}
    i=o_s+p,
    \qquad
    j=o_s+q .
\end{equation}
Therefore, the batch-level sparse multiplication satisfies
\begin{equation}
\label{eqns3-16}
    y_{o_s+p}
    =
    \sum_{q=1}^{N_s}
    A^{(s)}_{pq}
    \mathcal{N}^{(s)}_{\theta,q},
    \qquad
    p=1,2,\ldots,N_s .
\end{equation}
Equivalently, using the nonzero-entry representation,
\begin{equation}
\label{eqns3-17}
    y_i
    =
    \sum_{m:\,i_m=i}
    a_m
    \mathcal{N}^{\mathrm{bat}}_{\theta,j_m},
\end{equation}
where
\begin{equation}
\label{eqns3-18}
    i_m=o_s+p_m,
    \qquad
    j_m=o_s+q_m,
    \qquad
    a_m=A^{(s)}_{p_mq_m}.
\end{equation}
is mathematically equivalent to
\begin{equation}
\label{eqns3-19}
    y=A^{\mathrm{bat}}\mathcal{N}^{\mathrm{bat}}_\theta .
\end{equation}
The physics loss in FEMONet is therefore computed as
\begin{equation}
\label{eqns3-20}
    L^{\mathrm{bat}}(\theta)
    =
    \frac{1}{N_{\mathrm{bat}}}
    \sum_{i=1}^{N_{\mathrm{bat}}}
    \left|
    y_i-b^{\mathrm{bat}}_i
    \right|^2 .
\end{equation}
Substitute Eq.~(\ref{eqns3-17}) and Eq.~(\ref{eqns3-19}) into Eq.~(\ref{eqns3-20}),the above loss is equivalent to
\begin{equation}
    L^{\mathrm{bat}}(\theta)
    =
    \frac{1}{N_{\mathrm{bat}}}
    \sum_{i=1}^{N_{\mathrm{bat}}}
    \left|
    \sum_{j=1}^{N_{\mathrm{bat}}}
    A^{\mathrm{bat}}_{ij}
    \mathcal{N}^{\mathrm{bat}}_{\theta,j}
    -
    b^{\mathrm{bat}}_i
    \right|^2 .
\end{equation}

This formulation converts the evaluation of the physical constraint into a sparse vector operation. The computational cost of the residual evaluation scales with the number of nonzero entries,
\begin{equation}
    \mathcal{O}
    \left(
    \sum_{s=1}^{B}
    \mathrm{nnz}(A^{(s)})
    \right),
\end{equation}
rather than the cost of differentiating the neural-network output with respect to spatial coordinates at all quadrature points.

This is fundamentally different from VPINN. In VPINN, the neural network represents the field as a continuous coordinate-dependent function,
\begin{equation}
    v_\theta(\mathbf r)=\mathcal{N}_\theta(\mathbf r),
\end{equation}
and the weak-form residual is evaluated by directly inserting $v_\theta$ into the variational equation. For example, a typical weak-form residual contains terms such as
\begin{equation}
    \int_{\Omega}
    \nabla v_\theta
    \cdot
    \nabla w_i
    \,d\Omega ,
\end{equation}
or, for electromagnetic scattering problems,
\begin{equation}
    \int_{\Omega}
    \nabla\times v_\theta
    \cdot
    \overline{\mu}_r^{-1}
    \nabla\times w_i
    \,d\Omega ,
\end{equation}
where $w_i$ is a test function. In this case, the spatial derivatives of the neural-network output, such as
\begin{equation}
    \nabla v_\theta,
    \qquad
    \nabla\times v_\theta,
\end{equation}
must be computed through automatic differentiation.

If the neural network is expressed as a composition of $L$ layers,
\begin{equation}
    \mathcal{N}_\theta
    =
    \mathcal{N}_L
    \circ
    \mathcal{N}_{L-1}
    \circ
    \cdots
    \circ
    \mathcal{N}_1 ,
\end{equation}
then the spatial derivative is obtained by the chain rule,
\begin{equation}
    \frac{\partial v_\theta}{\partial x}
    =
    \frac{\partial \mathcal{N}_L}{\partial z_{L-1}}
    \frac{\partial \mathcal{N}_{L-1}}{\partial z_{L-2}}
    \cdots
    \frac{\partial \mathcal{N}_1}{\partial x}.
\end{equation}
During backpropagation, the gradient of the VPINN loss with respect to the trainable parameters further involves differentiating these spatial derivative terms with respect to $\theta$. Therefore, the computational graph contains both coordinate derivatives and parameter derivatives, which increases both computational cost and memory usage.

In contrast, FEMONet avoids coordinate-based automatic differentiation in the physics loss. The derivatives of finite-element basis functions are evaluated during FEM assembly and stored in the sparse matrix $A$. During training, the network only predicts the coefficient vector $\mathcal{N}_\theta$, and the physical residual is computed as
\begin{equation}
    A\mathcal{N}_\theta-b .
\end{equation}
The gradient of the loss follows the compact form derived as mentioned:
\begin{equation}
    \nabla_\theta L(\theta)
    =
    \frac{2}{N}
    \widetilde{J}(\theta)^T
    \widetilde{A}^T
    \left(
    \widetilde{A}
    \widetilde{\mathcal{N}}_\theta
    -
    \widetilde{b}
    \right),
\end{equation}
where $\widetilde{J}(\theta)$ is only the Jacobian of the network output with respect to the trainable parameters. No spatial derivative of the network output is required in the computational graph.

\section{Specific case settings}

This section summarizes the case-specific physical settings used to construct the operator-parameter sample sets. In all cases, the FEM matrices and load vectors were generated from parameterized COMSOL Multiphysics models through MATLAB scripts. Unless otherwise stated, $80\%$ of the generated samples were randomly selected for training and the remaining $20\%$ were used for testing. The FEM reference solutions were computed only for validation and were not used as supervised labels during FEMONet training. Additional implementation details are available in the public \href{https://github.com/HUST-CPO/FEMONet-Robust-and-universal-Operator-learning-for-optical-scattering-problem-via-MIONet}{GitHub repository}\cite{GitHub}

\subsection*{Basic lossless scatterers}

Five basic lossless scatterers shown in Fig.~\ref{fig3}(a) are considered, with geometric lengths ranging from $300$ to $650~\mathrm{nm}$ and a refractive index of $1.45$. The structures are placed in a $2~\mathrm{\mu m} \times 2~\mathrm{\mu m}$ square domain with scattering boundary condition (SBC) and excited by a TM-polarized plane wave $(\lambda = 1.55~\mathrm{\mu m}, E_0 = 1~\mathrm{V/m}, \theta = 0^{\circ})$. The FEM reference solutions of Eq.~(\ref{Ab}) are obtained using triangular elements with mesh size $\lambda/10$ and are used only for independent validation. The network architecture is DeepONet consists of Branch net 2 and the Trunk net in Fig.~\ref{fig2}, with $\varepsilon_r$ sampled on a $128 \times 128$ grid and $r_i$ denoting the vertex coordinates of scalar basis functions. The samples contains 1,558 unlabeled parametric samples, split into 1,246 training and 312 testing samples. Training is performed for 1,000 epochs using Adam with a cosine-annealed learning rate from $1\times10^{-3}$ to $1\times10^{-5}$

\subsection*{Single metallic scatterers}

This case is composed of seven metallic structures ($320-620~\mathrm{nm}$) based on QNM theory\cite{chen2020scattering,chen2021arbitrary}, as shown in Fig.~\ref{fig4}(a). The model employs the MIONet architecture shown in Fig.~\ref{fig2}, where the input $\lambda$ is the normalized wavelength, $\varepsilon_r$ and $E^i$ are processed as $128 \times 128$ matrices, while $r_i$ denotes the basis function vertex coordinates. Gold (Au) properties are determined via the Drude model across $1.5~\mathrm{\mu m}-2 ~\mathrm{\mu m}$. Excitation involves a TM-polarized plane wave ($|E|=1 V/m$) at incident angles $\theta\in[0^{\circ},315^{\circ}]$ with a $45^{\circ}$ step. The computation domain, boundary conditions, mesh length, sample preparation, and training settings are consistent with those in Case 1. A total of 50,688 unlabeled parametric space samples are generated and randomly divided into 40,550 training samples and 10,138 testing samples.

\subsection*{Multiple metallic scatterers}

This case is composed of arrayed metallic scatterers. As illustrated in Fig. \ref{fig5}(a), the configuration features a $1\times3$ array composed of the third and fifth scatterer structures selected from Fig. \ref{fig4}(a). The scatterers undergo random rotations ($\theta\in[0^{\circ},360^{\circ}]$) and scaling (0.8–1.2 times) around their centers, with a $1~\mathrm{\mu m}$ adjacent spacing. The computational domain is a $4~\mathrm{\mu m}$ square. The network architecture consists of Branch net 2,3 and the Trunk net in Fig.~\ref{fig2}, where the inputs $\varepsilon_r$ and $E^i$ are processed as $128 \times 128$ matrices, while $r_i$ denotes the basis function vertex coordinates. The boundary conditions, incident field, mesh size, sample preparation, and training settings are consistent with those in the previous cases. Of the 3,456 parameter samples generated, 2,765 were used for training and 691 for testing.

\subsection*{Plasmonic nanostructure}

This designated nanostructure is derived from a literature-reported device configuration\cite{jiang2020temperature}, which has been comprehensively validated through both numerical simulations and experimental measurements. 

As illustrated in Fig.~\ref{fig7}(a), the NWoM structure consists of an Au nanowire with a pentagonal cross-section $(d = 62~\mathrm{nm}, r = 4~\mathrm{nm})$ placed on a $h = 200~\mathrm{nm}$ Au mirror, separated by a $g = 5~\mathrm{nm}$ $\mathrm{Al}_2\mathrm{O}_3$ spacer and covered by a conformal $t = 5~\mathrm{nm}$ $\mathrm{Al}_2\mathrm{O}_3$ capping layer. The structure is excited by boundary condition of SBC with a normally incident light $|E|=1~\mathrm{V/m}$ and polarization angle $\theta = 0^{\circ}$. Au dispersion over $500$–$800~\mathrm{nm}$ follows the Drude model, while the mesh size, sample preparation, and training settings remain consistent with the previous cases. The network architecture consists of Branch net 1,2 and the Trunk net in Fig.~\ref{fig2}, where the input $\lambda$ is the normalized wavelength, $\varepsilon_r$ is processed as $128 \times 128$ matrices, while $\boldsymbol{r}_i$ denotes the endpoint-coordinate descriptor associated with the $i^{th}$ vector-basis degree of freedom. The computational samples holds 41 configurations ($5~\mathrm{nm}$ interval per sample), split into 33 training and 8 testing samples.

\subsection*{Three-dimensional metasurface}

This case study focuses on a three-dimensional (3D) metasurface unit cell structure, which is derived from a literature-reported device configuration\cite{fan2017visible}. The unit cell as shown in Fig.~\ref{fig8}(a), possesses a period of $P_x = P_y = 250~\mathrm{nm}$ and consists of $\text{TiO}_2$ nanofins ($L = 190~\mathrm{nm}$, $W = 90~\mathrm{nm}$, $h_1 = 600~\mathrm{nm}$, $\theta = 0^{\circ}$) on an Ag mirror ($h_2 = 150~\mathrm{nm}$). Material dispersion over $540–800~\mathrm{nm}$ is defined by the Devore ($\text{TiO}_2$), Schinke (Si), and Ciesielski (Ag) models. The network architecture consists of Branch net 1,2 and the Trunk net in Fig.~\ref{fig2}, where the input $\lambda$ is the normalized wavelength, $\varepsilon_r$ is processed as $128 \times 128 \times 128 $ matrices, while $r_i$ denotes the vector basis function center coordinates. The computational samples holds 261 configurations, split into 209 training and 52 testing samples.

\section{Gaussian fitting parameters}
\begin{table}[h]
\centering
\caption{Mean $\mu$ and standard deviation $\sigma$ of log$_{10}$ MSE from Gaussian fitted curves.}
\label{tab:mse_gaussian_fits}
\begin{tabular}{c c c c}
\toprule
\textbf{Sample} & \textbf{Process} 
& $\boldsymbol{\mu_{\log_{10}(\mathrm{MSE})}}$ 
& $\boldsymbol{\sigma_{\log_{10}(\mathrm{MSE})}}$ \\
\midrule
\multirow{4}{*}{1} & V-Train & -3.67 & 0.59 \\
                   & V-Test  & -3.65 & 0.65 \\
                   & Train  & -4.85 & 0.60 \\
                   & Test  & -4.77 & 0.71 \\
\midrule
\multirow{2}{*}{2} & Train & -2.97 & 0.18 \\
                   & Test  & -2.93 & 0.20 \\
\midrule
\multirow{2}{*}{3} & Train & -5.08 & 0.11 \\
                   & Test  & -4.83 & 0.16 \\
\midrule
\multirow{2}{*}{4} & Train & -3.87 & 0.21 \\
                   & Test  & -3.84 & 0.24 \\
\midrule
\multirow{2}{*}{5} & Train & -4.95 & 0.46 \\
                   & Test  & -5.08 & 0.39 \\
\midrule
\multirow{2}{*}{S} & Train & -1.90 & 0.27 \\
                   & Test  & -1.63 & 0.45 \\
\bottomrule
\end{tabular}
\end{table}
\section{Comparative ablation study and unseen samples supplementation}
\begin{figure}
    \centering
    \includegraphics[width=16cm]{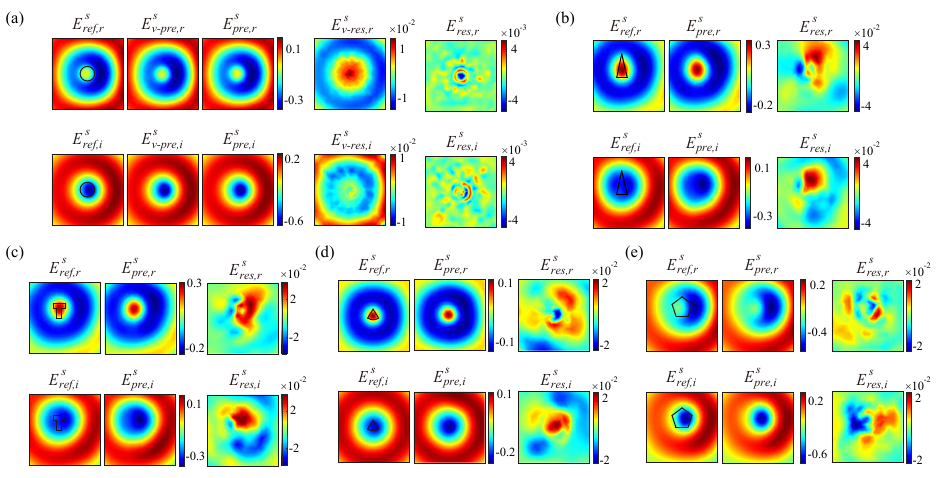}
    \caption{(a) Scattering field distributions for another representative lossless scatterer. Columns (left to right): FEM reference $E_{ref}^s$, VPINN prediction $E_{v-pre}^s$, FEMONet prediction $E_{pre}^s$, and their respective errors $E_{v-res}^s$ and $E_{res}^s$. (b-e) Extrapolation capabilities of FEMONet evaluated on unseen scatterer geometries (left to right: FEM reference $E_{ref}^s$, FEMONet prediction $E_{pre}^s$, and errors $E_{res}^s$). Rows indicate the real and imaginary parts.}
    \label{figs1}
\end{figure}
Figs.~\ref{figs1}(a) visualize the electric fields predicted by FEMONet for another representative sample randomly selected from the test set. Similarly to the representative sample in the main text, the residual of FEMONet is significantly lower than the enhanced VPINN, and the maximum errors of FEMONet are localized within specific regions, in contrast to the globalized error distribution exhibited by VPINN. In addition to the prediction result of the semi-circular scatterer presented in the main text, the results of the other four unseen samples are shown in Fig.~\ref{figs1}(b-e). FEMONet maintains high fidelity to the ground truth. 

\section{Sparse-sample supplementation experiment}
\begin{figure}
    \centering
    \includegraphics[width=14cm]{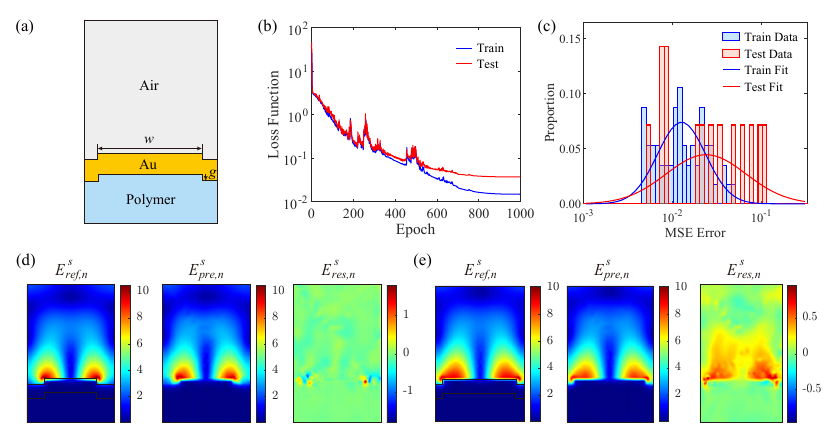}
    \caption{(a) Geometric structures of the scatterers. (b) Loss convergence curves. (c) MSE histograms with corresponding Gaussian fits. (d, e) Electric field norm distribution. Columns (left to right): FEM reference $E_{ref}^s$, FEMONet prediction $E_{pre}^s$ and error $E_{res}^s$.}
    \label{figs2}
\end{figure}
As illustrated in Fig.~\ref{figs2}(a), the structure has a period of $920~\mathrm{nm}$, a groove depth of $g=55 ~\mathrm{nm}$, and a duty cycle of $\mathrm{w}=0.2-0.9$, defined as the ratio between the ridge width and the period. the polymer grating had a thickness of $300~\mathrm{nm}$ and a refractive index of 1.55. The grating was coated with a $150~\mathrm{nm}$ gold layer. The dispersion of gold was modeled by the Brendel–Bormann model from the material library in COMSOL Multiphysics. We considered a air layer above the gold layer. Since grating is a periodic structure, we modeled it by using periodic condition along the direction of periodicity. The grating structure was illuminated from the top of the model window at a normal incidence. The illumination was a linearly transverse magnetic (TM) polarized plane wave, while the mesh size, sample preparation, and training settings remain consistent with the previous cases. The network architecture consists of Branch net 2 and the Trunk net in Fig.~\ref{fig2}, where the input $\varepsilon_r$ is processed as $128 \times 128$ matrices, while $r_i$ denotes the vector basis function center coordinates. The computational samples has 71 configurations ($\Delta\mathrm{w}=0.01$ interval per sample), split into 57 training samples and 14 testing samples.

Fig.~\ref{figs2}(b) illustrates the loss convergence curves, with training and testing error stabilizing at $1.5 \times 10^{-2}$ and $3.7 \times 10^{-2}$, respectively. The corresponding MSE Gaussian fits (Fig.~\ref{figs2}(c)) confirms that the model’s good generalization ability under sparse sample settings is not an isolated case. Fig.~\ref{figs2}(d,e) shows the electric field norm distributions predicted by FEMONet at $\mathrm{w}=0.6$ and $\mathrm{w}=0.83$. This case demonstrates that FEMONet can simulate surface plasmon polariton in plasmonic nanostructures using sparse samples.

\bibliographystyle{unsrt}  
\bibliography{references}

\end{document}